\documentclass[acmtog,nonacm]{acmart}

\copyrightyear{2026}
\acmYear{2026}
\acmConference[SIGGRAPH Conference Papers '26]{Special Interest Group on Computer Graphics and Interactive Techniques Conference Conference Papers}{July 19--23, 2026}{Los Angeles, CA, USA}
\acmBooktitle{Special Interest Group on Computer Graphics and Interactive Techniques Conference Conference Papers (SIGGRAPH Conference Papers '26), July 19--23, 2026, Los Angeles, CA, USA}
\acmDOI{10.1145/3799902.3811166}
\acmISBN{979-8-4007-2554-8/2026/07}

\acmSubmissionID{959}

\usepackage{booktabs} %
\usepackage{xspace}
\usepackage{listings}
\usepackage{amsmath}
\usepackage{amsfonts}
\usepackage{cleveref}
\usepackage{microtype}
\usepackage{enumitem}
\usepackage{multirow}
\usepackage{overpic}
\usepackage{colortbl}
\usepackage{tablefootnote}
\usepackage{hyperref}
\usepackage{caption}
\usepackage{subcaption}
\usepackage{cuted}
\hypersetup{
    colorlinks=true,
    linkcolor=blue,
    filecolor=magenta,      
    urlcolor=cyan,
    pdfpagemode=FullScreen,
}

\setlist{topsep=2pt, leftmargin=*}

\graphicspath{ {./images/} }

\def \eg {{\emph{e.g.},\thinspace}}
\def \ie {{\emph{i.e.},\thinspace}}

\newcommand{\name}{\emph{B-repLer}\xspace}
\newcommand{\dname}{\emph{BrepEDIT-240K}\xspace}

\newcommand{\papertitle}{\name: Language-guided Editing of CAD Models}

\makeatletter
\newcommand{\makesupplementtitle}{%
	\begin{strip}
		\centering
		{\@titlefont \papertitle\\ Supplementary Materials\par}
	\end{strip}%
}
\makeatother

\definecolor{green}{rgb}{0, 0.5, 0}
\definecolor{orange}{rgb}{0.6, 0.3, 0.1}
\definecolor{red}{rgb}{1.0, 0.0, 0.0}
\definecolor{teal}{rgb}{0.0, 0.4, 0.4}
\definecolor{purple}{rgb}{0.65,0,0.65}
\definecolor{saffron}{rgb}{0.95,0.75,0.2}
\definecolor{turquoise}{rgb}{0.0,0.4,0.8}
\definecolor{brown}{rgb}{0.5, 0.16, 0.16}
\definecolor{brickred}{rgb}{.6, .2 .1}
\definecolor{coral}{rgb}{1,0.45,0.33}
\definecolor{llightgray}{RGB}{230,230,230}

\newcommand{\rev}[1]{{{#1}}}

\author{Yilin Liu}
\affiliation{
  \institution{University College London}
  \city{London}
  \country{United Kingdom}}
\authornote{Corresponding author: Yilin Liu (whatsevenlyl@gmail.com)}

\author{Niladri Shekhar Dutt}
\affiliation{
  \institution{University College London}
  \city{London}
  \country{United Kingdom}
}

\author{Changjian Li}
\affiliation{
  \institution{University of Edinburgh}
  \city{Edinburgh}
  \country{United Kingdom}
}

\author{Niloy J. Mitra}
\affiliation{
  \institution{University College London}
  \city{London}
  \country{United Kingdom}
}
\affiliation{
  \institution{Adobe Research}
  \city{London}
  \country{United Kingdom}
}

\begin{document}
\title{\name: Language-guided Editing of CAD Models}

\begin{teaserfigure}
    \begin{overpic}[width=\textwidth]{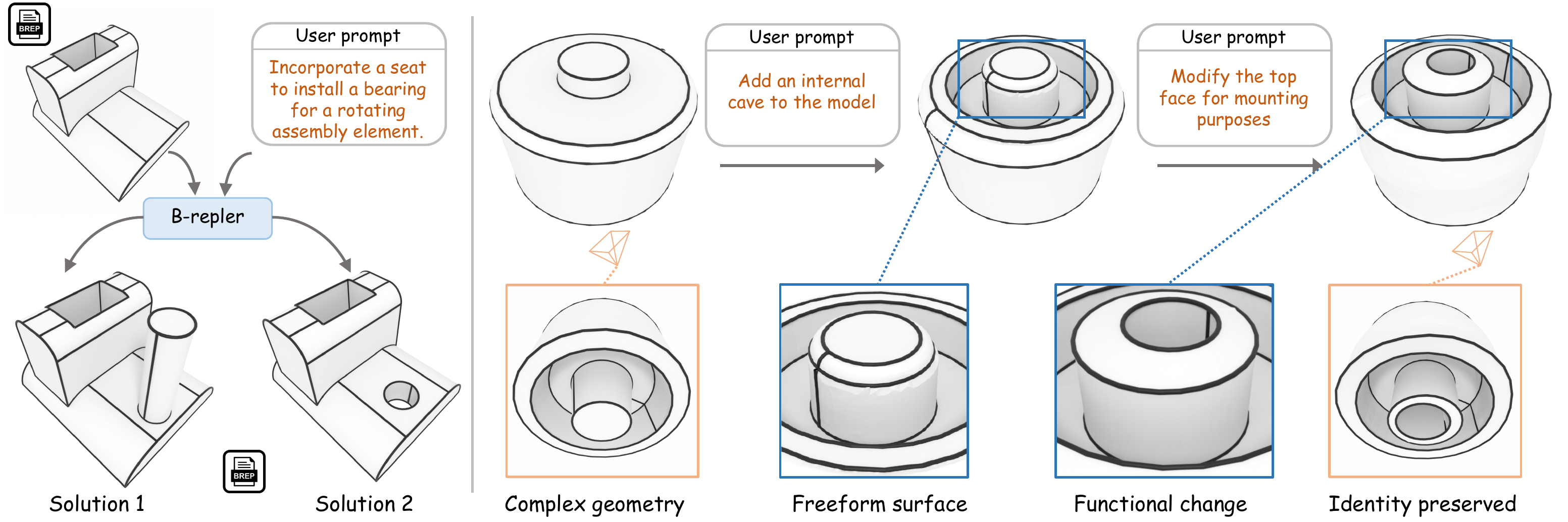} %
    \end{overpic}    
    \captionof{figure}{\textbf{Text-driven B-rep editing.} 
    We present \textit{\textbf{B-rep L}}atent \textit{\textbf{E}}dito\textit{\textbf{r}} (\name), the first framework to perform generative CAD editing directly in the Boundary Representation (B-rep), 
    without the need for the construction history. Our method can interpret high-level user instructions (in orange) to perform fine-grained modifications on B-rep models with complex and freeform geometries. 
    By operating in the B-rep latent space, \name generates a diverse set of valid B-rep edits from a single fixed input (left).
    It also supports multi-edit for complex freeform surfaces and functional descriptions, while preserving identities in unchanged areas (right).
    }
    \label{fig:teaser}
\end{teaserfigure}

\begin{abstract}
Computer-Aided Design~(CAD) models, given their compactness and precision, remain the industry standard for designing and fabricating engineering objects. However, language-guided CAD editing is still in its infancy, largely due to missing semantic connection between user commands and underlying shape geometry, a problem exacerbated by the shortage of paired text-and-edit CAD datasets. While recent Multimodal Large Language Models (mLLMs) have attempted to bridge this gap, their reliance on CAD construction history --often an expensive and hard to obtain input-- severely limits their expressiveness and restricts their usage.
We present \name, a novel framework that directly connects natural language with editing CAD models by operating in a learned latent space. Importantly, our approach bypasses the need for construction history, enabling semantic edits on a wide range of geometries, from simple prismatic parts to complex freeform shapes defined by B-Spline surfaces. To facilitate this research, we introduce \dname, the first large-scale dataset for this task. We demonstrate how this paired dataset can be automatically generated, (user) validated, and scaled by leveraging existing CAD tools, in conjunction with mLLMs, to create the required paired data without relying on any external annotations. Our results demonstrate that B-repLer can accurately perform complex edits on complex CAD shapes, even when the input edit specifications are high-level and ambiguous to interpret, consistently producing valid, high-quality CAD outputs enabling a class of text-guided edits not previously possible.
Project page is at \url{https://yilinliu77.github.io/brepler.github.io/}.

\end{abstract}

\ccsdesc[500]{Computing methodologies~Shape modeling}
\ccsdesc[500]{Computing methodologies~Neural networks}
\ccsdesc[500]{Applied computing~Computer-aided design}

\keywords{Boundary representation; Shape Editing; Generative models}

\maketitle

\section{Introduction}
\label{sec:intro}

Computer-Aided Design (CAD) models are the definitive industry standard for engineering objects, prized for their ability to compactly and precisely encode complex objects. While its parametric nature supports methodical editing within CAD software, performing high-level and semantic edits, driven by natural language, remains an unsolved and difficult problem. The challenge is threefold: (i)~a \emph{semantic misalignment} between abstract user intent and required low-level geometric operations; (ii)~a \emph{data limitation} due to the absence of public CAD editing datasets with or without operations history; and the (iii)~\emph{fragile} nature of the CAD structure itself. In this work, we focus on Boundary Representation (B-rep) that encodes CAD object as a structured collection of parametric surfaces (\eg planes and B-Splines), trimmed by curves and joined at vertices, all governed by strict topological rules; this complexity, however, means minor mismatches/errors can easily result in invalid models.

\begin{figure}[t!]
    \centering
    \begin{overpic}[width=0.99\linewidth]{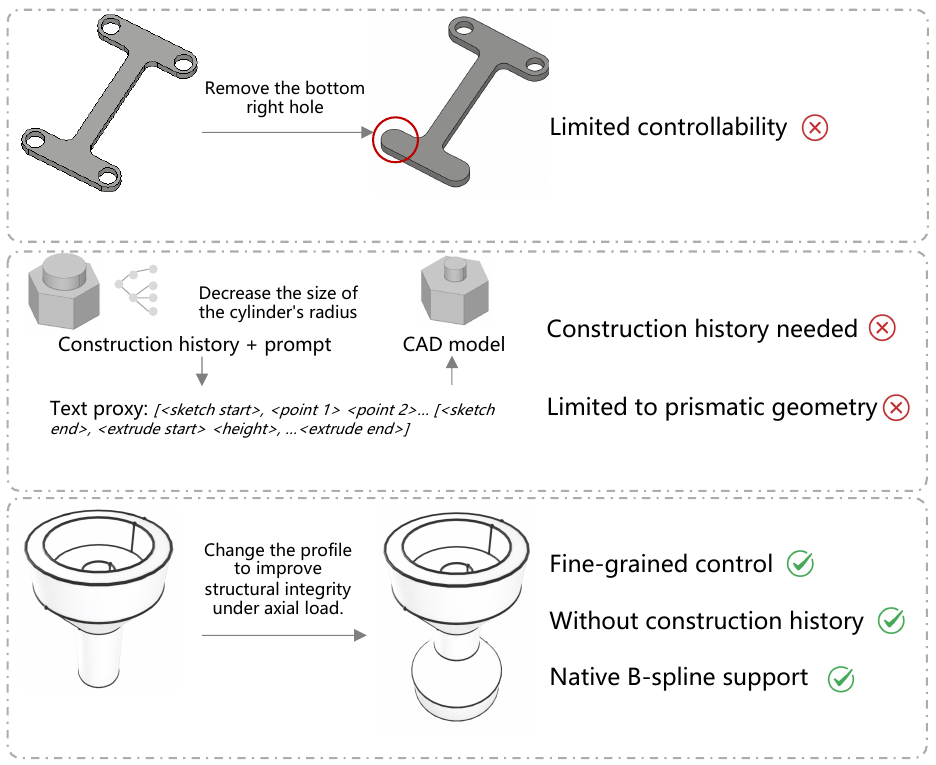} %
        \put(16,58) {\footnotesize (a) mLLM image editing + img2brep baseline~\cite{HolaBRep25}}
        \put(32,30.5) {\footnotesize (b) Finetuned mLLM~\cite{yuan2025cadeditor}}
        \put(38,1.5) {\footnotesize (c) Our method}
    \end{overpic}
    \caption{\textbf{B-rep Editing.} Text-based B-rep editing is challenging. For example, (a) a training-free pipeline usually involves multiple stages, where errors can occur and accumulate across different stages, limiting its controllability; (b) a trained mLLM~\cite{yuan2025cadeditor} used the expensive construction history as a text proxy to generate modified B-rep, while having limited shape and edit complexity; (c) we propose a novel \emph{native} B-rep editing approach, where we link the text embedding to B-rep \textit{latent}, thus enabling semantic-aware and fine-grained control over various primitives. 
    }
    \label{fig:motivation}
\end{figure}

Recently, in other domains (\eg images, videos), Multimodal Large Language Models (mLLMs) have achieved remarkable success in both synthesis and editing~\cite{oasis,dalle3,mgie,dutt2025monetgpt}. However, their success has not been translated to the CAD domain, where the aforementioned challenges create a barrier. Even specialized, training-based approaches that attempt to represent CAD operations as text~\cite{yuan2025cadeditor} are fundamentally constrained. For example, the recent CAD-Editor~\cite{yuan2025cadeditor} relies on a model's construction history, an expensive and often unavailable input\footnote{Current public data containing construction history is significantly simpler and 5--10 times less than datasets containing only the final B-rep.}, which limits its expressiveness to simple prismatic shapes; such construction-history based methods cannot handle the vast number of B-rep models that exist only as final boundary descriptions, as shown in \Cref{fig:motivation}. A truly general solution has remained elusive.

We propose a new approach, enabled by recent advances in direct B-rep autoencoders~\cite{brepgen24, HolaBRep25}. These models learn to map complex B-rep geometries into a continuous latent space. Our key insight is that this latent representation offers a solution to the core problems of B-rep editing. By operating in this space, we avoid brittle text proxies for geometry, allowing us to natively support complex, B-Spline-based shapes without their construction history. Furthermore, because the decoder is trained to produce valid outputs, this approach circumvents the fragility problem by design.

We introduce \name, which addresses the editing task by employing a generative variational autoregressive transformer to execute edits directly in the latent space. 
This architecture is specifically designed to accommodate the one-to-many nature of creative editing and the variable lengths of the resulting geometry sequences, while simultaneously eliminating quantization errors that otherwise compromise the precision inherent to the B-rep format. 
To overcome the challenge of data scarcity, we further introduce \dname, the first large-scale dataset established for this task.
We demonstrate that our automated generation process is essential not only for feasibility but also for ensuring the scalability of our methodology.

Extensive experiments demonstrate that our dataset features rich annotations, complex shapes, and diverse edit operations, while maintaining practical real-world utility. Furthermore, our proposed algorithm outperforms state-of-the-art methods while requiring significantly less information; notably, it is the first approach capable of performing B-rep editing based on high-level semantic and functional descriptions without relying on construction history, see \Cref{fig:teaser}.
To summarize, our main contributions include:

\begin{itemize}
    \item \name, the first text-driven B-rep model editing approach that interprets user instructions and performs edits in the B-rep latent space, enabling native and fine-grained B-rep editing without any modality conversion; 
    \item \dname, the first dataset that collects text-driven edits on complex B-rep models with freeform surfaces; 
    \item a variational autoregressive model that efficiently models B-rep latent sequences without the need for vector quantization, and adheres to text-guidance for editing. 
\end{itemize}

\section{Related Works}

\subsection{CAD Generation}
\label{sec:cad_generation}

Learning to generate CAD models has become a popular topic due to the compactness and expressiveness of the CAD representation.
\rev{
Based on early representation learning of CAD models~\cite{uvnet21,brepnet23,Cao20,AutoMate21,JoinABLe22,HGCAD23}, researchers have proposed various methods to generate CAD models in both sketch-and-extrude~\cite{deepcad21,you2024img2cad, LiG0Y23, XuPCWR21,RenZCLZ22, XuJLWF23, CADParser23, chen2024img2cad} and more general boundary representation (B-rep)~\cite{HolaBRep25,solidgen23,brepgen24,BrepDiff,DTGBrepGen}, using either diffusion or autoregressive models. Beyond generation, prior B-rep understanding work has also studied entity-level correspondence across CAD systems~\cite{BrepMatching}, showing that learning directly on B-rep topology enables robust matching without relying on proprietary persistent naming schemes.
While earlier approaches emphasized sequential generation, recent works~\cite{solidgen23,brepgen24,BrepDiff,HolaBRep25,DTGBrepGen} demonstrate that directly generating B-rep models leads to greater efficiency and expressiveness, especially for complex topologies and freeform surfaces. In particular, BrepDiff~\cite{BrepDiff} proposes a single-stage diffusion transformer over masked UV-grid face representations, avoiding the cascaded topology-then-geometry design used in prior multi-stage pipelines.
}
Our work leverages the CAD latent code from HoLa-BRep~\cite{HolaBRep25} ensuring a compact latent space.

In conditional generation, CAD reconstruction from point clouds, referred to as reverse engineering, has been explored through segmentation~\cite{parsenet20,primitivenet,hpnet21,bpnet23,sed23}, primitive fitting~\cite{complexgen22}, and mesh reconstruction~\cite{nvd24}. Owing to their 3D nature, these methods typically outperform alternatives based on image~\cite{you2024img2cad,chen2024img2cad,HolaBRep25},  text~\cite{khan2024textcad,HolaBRep25}, or mixed~\cite{xu2024cadmllm}.

\subsection{LLMs for xD Generation and Editing}
\label{sec:llm_3d}

\paragraph{LLMs in xD generation.}
Multimodal LLMs have been successfully used in various domain-specific tasks -- image editing~\cite{fu2023guiding,dutt2025monetgpt}, layout generation~\cite{littlefair2025flairgpt}, mathematical reasoning~\cite{romera2024mathematical}; this has sparked interest in utilizing them for 3D mesh and CAD generation. LLaMA-Mesh~\cite{wang2024llama} fine-tunes LLaMA~\cite{grattafiori2024llama} for 3D mesh generation by supervising it to directly generate vertex positions and faces. Other works adapt mLLMs to domain-specific 3D structures, including garments~\cite{bian2024chatgarment}, human pose and interaction~\cite{lin2024chathuman}, and motion~\cite{jiang2023motiongpt}.

\paragraph{LLMs in CAD generation.}
Recent works in CAD generation from natural language instructions and/or images have focused on fine-tuning mLLMs to produce CAD program sequences ~\cite{wang2025cad,yuan2025cadeditor,xu2024cadmllm}. CAD-GPT~\cite{wang2025cad} maps 3D positions and sketch plane rotation angles to a language feature space to achieve precise spatial localization in both image-to-CAD and text-to-CAD.
CADCodeVerify~\cite{alrashedy2025generating} directly prompts a pre-trained VLLM to write the CAD program and iteratively verify and improve the designed output from a CAD kernel.
CAD-MLLM~\cite{xu2024cadmllm} enables unified conditional generation from text, images, and point clouds by fine-tuning a pre-trained LLM to align features with CAD command sequences.
CAD-LlamaINS~\cite{li2025cad} presents a hierarchical annotation pipeline to translate the CAD sketch-extrude sequences into their code-like Structured Parametric CAD Code format with hierarchical semantic descriptions and fine-tune Llama~\cite{grattafiori2024llama} to generate parametric CAD models. 

\paragraph{Structured shape editing.}
Early shape editing methods, such as iWire \cite{iwire}, impose geometric constraints to preserve sharp features during interactive editing; 
Wei et al.~\cite{wei2020learning} align input shapes to template models, transferring edits via a shared latent space. 
However, editing CAD or general 3D models remains challenging due to limited paired training data. 
Sketch2CAD~\cite{Sketch2CAD} maps 2D sketch edits to 3D operations with a dataset of sketch-and-extrude sequences but remains bound to pre-defined, sequential construction steps. 
MeshPad~\cite{MeshPad} and FlexCAD~\cite{zhang2024flexcad} synthesize edits by masking or cropping mesh/CAD elements, training models to predict local modifications. 
While FlexCAD leverages LLMs to enable some level of primitive-level control (\eg altering a cylinder), it lacks the capacity for fine-grained semantic edits (\eg making the cylinder wider).
On the other hand, CADTalk~\cite{yuan2024cadtalk} tries to augment the CAD program with fine-grained part-wise semantic labels by combining program parsing with visual-semantic analysis using the recent foundational language and vision models. Their semantic labels make subsequent semantic-based CAD shape editing possible.
Along this direction, given a part-segmented 3D mesh and an editing text, ParSEL~\cite{ganeshan2024parsel} produces a parameterized editing program by leveraging LLMs and their novel analytical edit propagation. The editing is efficiently achieved in the parametric domain; however, they can only handle limited shape structures and editing requests expressed as affine transformations without changing the original topology. 

The \emph{concurrent} work, CAD-Editor~\cite{yuan2025cadeditor}, uses multimodal LLMs to generate synthetic edit triplets and fine-tunes an LLM to locate and modify specific CAD tokens.
Though effective, it remains constrained by the sketch-and-extrude format, limiting scalability to complex geometries (\eg B-spline primitives) and datasets lacking construction history (\eg ABC~\cite{ABC}). 
In contrast, our method operates directly on B-rep representations rather than procedural sequences. 

\begin{figure*}[!t]
    \centering
    \includegraphics[width=0.99\linewidth]{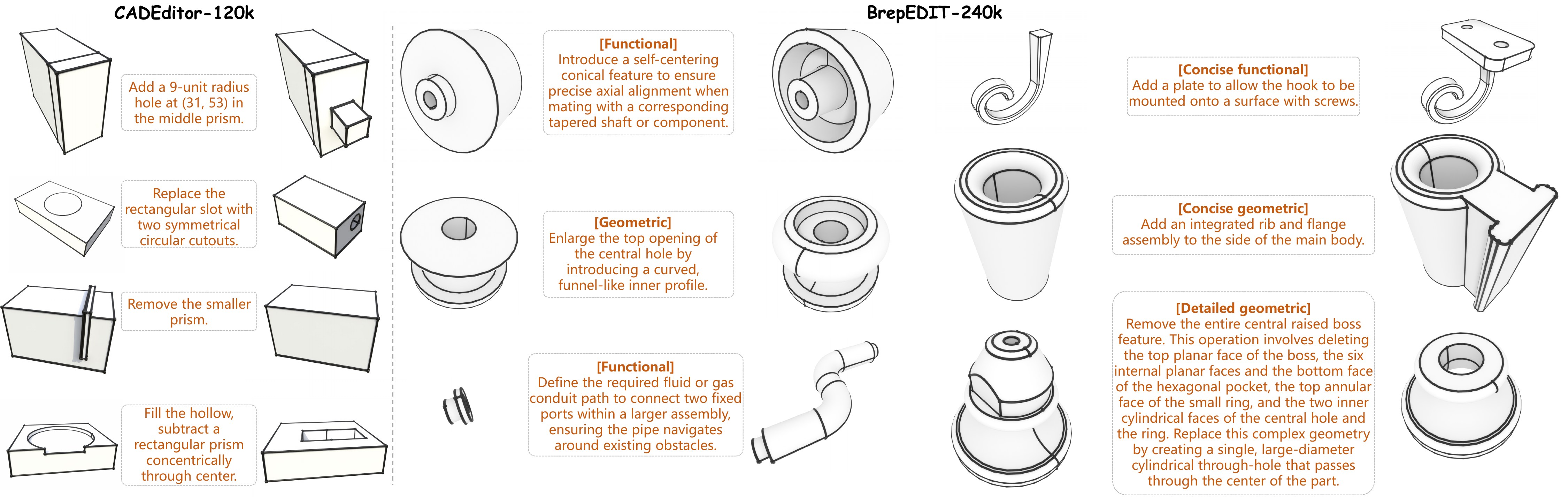}
    \caption{\textbf{Dataset comparison.}
    \dname has rich annotations across complex and practical CAD models, where CAD-Editor~\cite{yuan2025cadeditor} is intrinsically restricted by the sketch-and-extrude definition and limited annotation quality (zoom recommended).
    }
    \label{fig:res_dataset}
\end{figure*}

\section{\dname Dataset}
\label{sec:dataset}

\subsection{Overview}
\dname is built upon the ABC dataset~\cite{ABC}, where we enhance the existing B-rep models with semantic annotations that describe the editing instructions.
For each data point, we provide a pair of B-rep models: the pre-edit (before) model $M_b$ and the post-edit (after) model $M_a$ along with a set of varied text annotations $\{\hat{T}_{a\rightarrow b}, \hat{T}_{b\rightarrow a}\}_{i=1}^{K}$ that describe the editing instruction from $M_a \rightarrow M_b$ and their inversion. 
We also include the corresponding 2D bounding box $B \in \mathbb{R}^4$ of the edited region, which serves as a coarse localization cue for the modification, and the rendered images $I_b$ and $I_a$ of the original and edited models, respectively.
The bounding boxes and the images are associated with corresponding viewpoints (\ie edits are visible from the viewpoints, as explained later).
Following previous works~\cite{HolaBRep25,brepgen24}, we filter out simple shapes which have fewer than 10 faces to ensure sufficient geometric complexity, and shapes that fail to be encoded by the HoLa-BRep~\cite{HolaBRep25} encoder.
Additionally, we also conduct filtering and analysis on the editing operation to ensure the synthetic edits are valid and meaningful (detailed in Sec.~\ref{sec:dataset_construction}).
The resulting dataset covers 52k unique shapes, 240k addition/deletion operations, and the corresponding multi-level text annotations, see \Cref{fig:res_dataset}.

\begin{figure}[b!]
    \centering
    \begin{overpic}[width=0.99\linewidth]{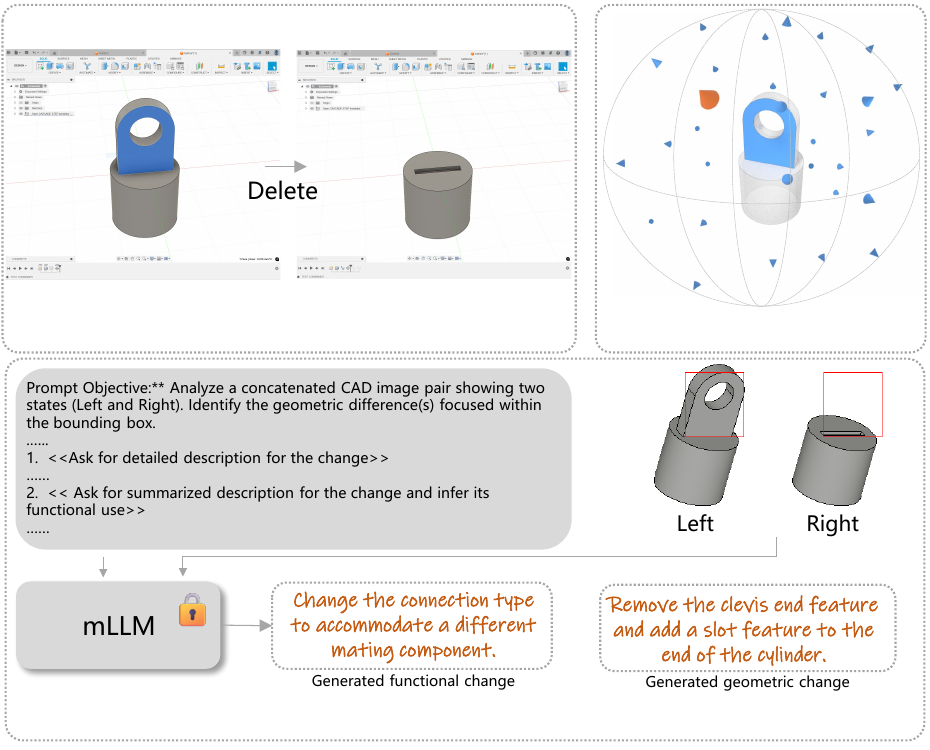} %
        \put(13,44) {\footnotesize (a) Face deletion in Fusion360}
        \put(67,44) {\footnotesize (b) Image Rendering}
        \put(22,2.5) {\footnotesize (c) Instruction labeling using an mLLM}
    \end{overpic}
    \caption{\textbf{Dataset construction}.
    We construct an annotated dataset in three main steps: 
    (a) Given a B-rep model, we randomly select an operating face, and we then employ \rev{Fusion360~\cite{fusion360}} to delete it. \rev{The geometric kernel of the CAD engine automatically "heals" the local topology and geometry to preserve the validity of the resulting B-rep model.}
    (b) For each successful deletion, we select the best viewpoint to render images before and after the operation, and 
    (c) combine them with additional ``Left'' and ``Right'' texts to form an input image sent to an mLLM (\ie Gemini 2.5 Pro~\cite{gemini}) to produce the corresponding editing instructions.
    }
    \label{fig:dataset_operation}
\end{figure}

\begin{figure*}[t!]
    \centering
    \begin{overpic}[width=0.99\linewidth]{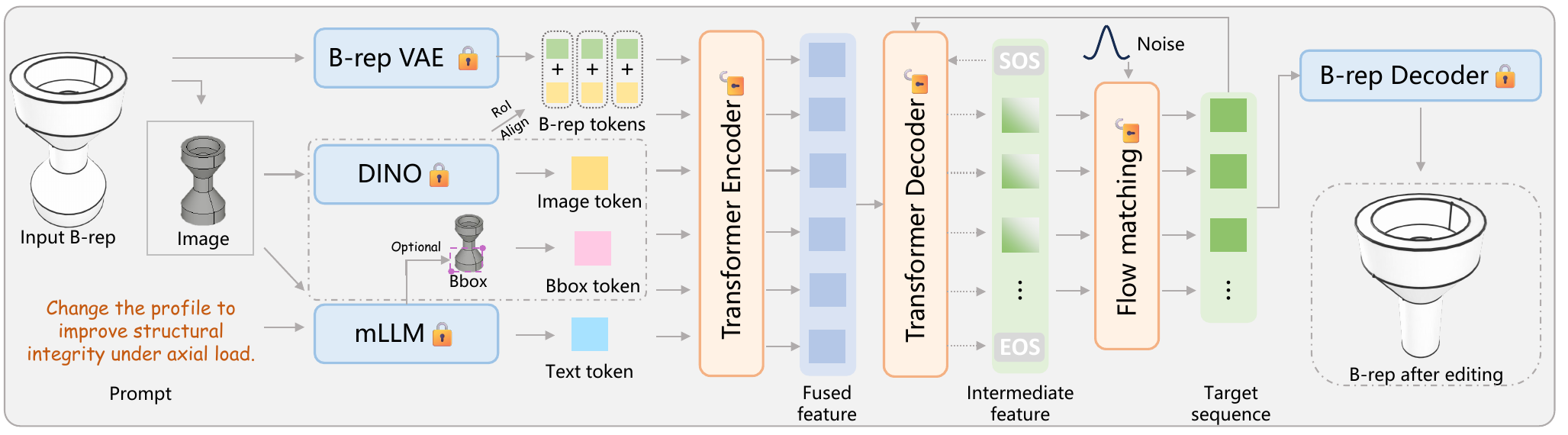} %
    \end{overpic}
    \caption{\textbf{Method  Overview.}
        We process the input B-rep model and user prompt using a pre-trained mLLM and a B-rep encoder to extract visual, textual, and B-rep features. These features are fused by a transformer encoder. Subsequently, a decoder autoregressively attends to the fused features to produce a target embedding, which conditions a flow matching network to generate the edited B-rep CAD model.
    }
    \label{fig:pipeline}
\end{figure*}

\subsection{Dataset Construction}
\label{sec:dataset_construction}
The key to our data construction is to obtain the paired B-rep models before and after editing. However, since there is no construction history in the ABC dataset, preparing paired models is challenging.
While prior works, like MeshPad~\cite{MeshPad}, demonstrate the feasibility of synthesizing editing instructions and pre-edit shapes, directly applying similar strategies to B-reps encounters challenges.
(i)~First, unlike triangle meshes, B-rep models are highly structured; edits must preserve topological and geometric integrity—random modifications, such as cropping, often lead to invalid or incomplete representations.
(ii)~Second, user-provided editing instructions are frequently high-level or abstract (\eg ``strengthen this desk'' or ``add a component for additional fastening''), requiring a semantic-aware generation pipeline capable of interpreting and executing such loosely-specified prompts (\eg what specific structural or geometric changes are needed to ``strengthen'' an object).

To address these challenges, we construct our dataset by leveraging the semantic reasoning capabilities of an mLLM (\ie Gemini 2.5 Pro\cite{gemini}) and the geometric robustness of a modern CAD engine (\ie Fusion360 \cite{fusion360}).
Specifically, given a B-rep model from ABC dataset~\cite{ABC}, we proceed as follows:
\begin{enumerate}
    \item \textit{Face deletion} (\Cref{fig:dataset_operation}(a)). We treat the input B-rep model as the \textit{post-edit} state $M_a$, and iterate over all its faces, attempting to apply the deletion operation to each face. 
    \rev{Then we rely on the healing mechanism of the CAD engine to check if the local topology and geometry can be adaptively adjusted to preserve the validity of the resulting B-rep model.}
    If the resulting model $M_b$ remains valid after the operation, we record the model pair $(M_b, M_a)$ as a valid edit instance.
    \rev{Due to the computation budget, we choose one case for each model where the edit involves the most changes to the geometry among all the valid edits.}

    \item \textit{Image rendering} (\Cref{fig:dataset_operation}(b)). For each successfully recorded model pair, we select the best view to render their images as $I_b$ and $I_a$. Here, the best view is chosen from a predefined 32 views that are uniformly sampled from a surrounding sphere, and it can best reveal point samples on the operational face. Additionally, we project the 3D points sampled from the selected face in the face deletion step back onto the rendered image and calculate their image axis-aligned 2D bounding box as $B$, recording its top-left and bottom-right corners. 
    
    \item \textit{Instruction labeling} 
    (\Cref{fig:dataset_operation}(c)). {
    Finally, we generate annotations by querying an mLLM to identify and infer the change between the two rendered images (\Cref{fig:dataset_operation}(c)). We provide the mLLM with the concatenated before/after images and the 2D bounding box of the changed area, prompting it to generate: (i) shape summary; (ii) detailed geometric description; (iii) coarse geometric description; (iv) detailed functional description; and (v) coarse functional description.
    Those annotations are requested in a sequential, detailed to concise manner to help the mLLM to better interpret the change.
    More details are shown in the supplementary.
    }

    Note that although we only apply the \texttt{Delete} operation, we obtain the \texttt{Add} in the inverse direction. We query the mLLM to obtain $\hat{T}_{a\rightarrow b}$ and $\hat{T}_{b\rightarrow a}$ in the same call. 
\end{enumerate}
\rev{This design also helps explain why the resulting supervision can generalize beyond literal delete/add instructions. Although the paired geometry is synthesized from face deletion and its inverse, the resulting shape changes can be substantial and are labeled semantically by the mLLM using operators such as replace, enlarge, or increase whenever appropriate. As a result, the model is trained on semantically enriched edit pairs rather than on a narrow set of literal operation names.} Also, using Fusion360 and Gemini ensures the resulting annotated edit is both structurally consistent and semantically plausible, enabling the generation of meaningful data for text-conditioned B-rep editing.

\section{Method}
\label{sec:method}

\paragraph{Overview.} 
Editing B-rep CAD models directly from high-level textual instructions presents a notable challenge due to the significant domain gap between abstract instructions and low-level geometric modifications. Furthermore, complex or abstract instructions often allow for multiple valid interpretations, leading to diverse potential modifications. As illustrated in \Cref{fig:pipeline}, \name addresses these challenges by framing the task as a sequence-to-sequence translation problem, utilizing a variational autoregressive transformer.

\paragraph{Encoding.} 
We adopt the holistic B-rep latent space proposed in~\cite{HolaBRep25} for its concise representation and ability to preserve CAD model validity. To this end, we first encode both $M_b$ and $M_a$ using the pretrained HoLa Encoder~\cite{HolaBRep25}.
As our autoregressive model is sensitive to sequence order, we first establish a canonical ordering by computing the center position of each surface and sorting them in ascending XYZ order.
This process yields the ordered latent representations $\{H_b^i\}_{i=1}^{N} \in \mathbb{R}^{N \times d}$ and $\{H_a^i\}_{i=1}^{N'} \in \mathbb{R}^{N' \times d}$. Here, $H_{b}^i$ and $H_{a}^i$ denote the latent vectors of the $i$-th face in $M_b$ and $M_a$, respectively, and $d=32$ is the dimensionality of each feature vector.

Given the text instruction $T_{b\rightarrow a}$, and a rendered image $I_b$ of $M_b$, we first extract multi-modal features to form the input sequence $S_b \in \mathbb{R}^{(N+3) \times 768}$ for the transformer.
\begin{itemize}
    \item The B-rep latent features before editing, $\{\mathcal{F}_\text{brep}^i\}_{i=1}^{N} \in \mathbb{R}^{N \times 768}$, mapped from the HoLa-encoded latent ${H_b}$ via a trainable projection: MLP$_{\text{brep}}: \mathbb{R}^{32} \rightarrow \mathbb{R}^{768}$.
    
    \item A text feature $\mathcal{F}_{\text{text}} \in \mathbb{R}^{1 \times 768}$, obtained via a frozen Qwen2.5 text encoder $\text{Qwen}: \textit{Text} \rightarrow \mathbb{R}^{2560}$ and a trainable projection: MLP$_{\text{text}}: \mathbb{R}^{2560} \rightarrow \mathbb{R}^{768}$.

    \item An image feature $\mathcal{F}_{\text{img}} \in \mathbb{R}^{1 \times 768}$, extracted using a frozen DINOv2 encoder DINO$:\mathbb{R}^{3 \times 224 \times 224} \rightarrow \mathbb{R}^{1024}$ followed by a trainable projection: MLP$_{\text{img}}: \mathbb{R}^{1024} \rightarrow \mathbb{R}^{768}$.

\end{itemize}

\rev{
A bounding box of the potentially edited region can be optionally provided as an additional input to the model, serving as a coarse localization cue for the modification.
It can be obtained either from an mLLM inference or from a user-provided input.
A bounding box feature $\mathcal{F}_{\text{bbox}} \in \mathbb{R}^{1 \times 768}$ is then obtained by passing the bounding box (\ie top-left and bottom-right corner points) through a trainable projection: MLP$_{\text{box}}: \mathbb{R}^4 \rightarrow \mathbb{R}^{768}$.
If the bounding box is not provided, we simply set $\mathcal{F}_{\text{bbox}}$ to a learnable embedding.
}

Although both 2D image features and 3D B-rep latent features are provided as input, they are not naturally aligned. To bridge this gap, we enhance each B-rep face feature with localized image context. Specifically, we extract the bounding box of each face in $M_b$ and use RoIAlign and an MLP to retrieve (crop) and transform the corresponding regional feature from the DINO feature map, encoded as $\{\mathcal{F}_\text{roi}^{i} \}_{i=1}^N \in \mathbb{R}^{N \times 768}$. This feature is then \textit{added} to the corresponding B-rep latent $\mathcal{F}_{\text{brep}}^i$ to inject image-aware information. 

The input sequence $S_b$ is passed through a standard transformer encoder $\mathcal{E}$, yielding a fused feature representation $\mathcal{F}_{\text{src}} \in \mathbb{R}^{(N+3) \times 768}$ that jointly captures image, text, and geometry semantics.
Specifically, 
$S_b := \{  \{\mathcal{F}_\text{brep}^i + \mathcal{F}_\text{roi}^{i}\}_{i=1}^{N}, 
\mathcal{F}_\text{img},
\mathcal{F}_\text{bbox},
\mathcal{F}_\text{text}
\}$.

\paragraph{Decoding.}
The goal of decoding is to predict the B-rep latent sequence $\{H_a^i\}_{i=1}^{N'} \in \mathbb{R}^{N' \times 32}$ corresponding to the edited model $M_a$. To address the one-to-many ambiguity inherent in B-rep editing, we propose an interleaved decoding process that combines the autoregressive transformer with a flow matching network~\cite{flowmaching2023}.
We distinguish the model's operation during training and inference.

\noindent \textbf{Training.} We employ teacher forcing for efficient, parallel computation:
\begin{itemize}
    \item \textit{Transformer Input}: The transformer decoder takes as input the entire ground-truth B-rep latent sequence $\{H_a^t\}_{t=1}^{N'+1}$, where a start-of-sequence (SOS) token is padded. These $\mathbb{R}^{32}$ tokens are first projected to $\mathbb{R}^{768}$ feature space.
    \item \textit{Condition Prediction}: In a single forward pass, the decoder attends to the fused context $\mathcal{F}_{\text{src}}$ and the ground-truth history to predict the full sequence of intermediate conditioning features $\{\mathcal{F}_\text{inter}^t\}_{t=1}^{N'+1} \in \mathbb{R}^{(N'+1) \times 768}$.
    \item \textit{EOS classification:} A classifier head is trained on the last feature from $\{\mathcal{F}_\text{inter}^t\}$ to predict whether the current step is the end of the sequence.
    \item \textit{Flow Matching Loss}: The FM model $(\mathbb{R}^{N' \times 32},\mathbb{R}^{N' \times 768}) \rightarrow \mathbb{R}^{N' \times 32}$, is then trained to reconstruct the ground-truth tokens $\{H_a^t\}_{t=1}^{N'}$ from Gaussian noise $G \in \mathbb{R}^{N' \times 32}$, conditioned on the intermediate features $\{\mathcal{F}_\text{inter}^t \in \mathbb{R}^{N' \times 768}\}$. This optimization follows the standard flow matching objective.
\end{itemize}

\noindent \textbf{Inference.} The model generates the output sequence one token at a time. For each autoregressive step $t$ (where $t=1, 2, \ldots, N'$):
\begin{itemize}
    \item \emph{Condition Prediction}: The transformer decoder attends to the fused context $\mathcal{F}_{\text{src}}$ and the previously generated tokens $\{H_a^0, ..., H_a^{t-1}\}$ (projected to $\mathbb{R}^{768}$) to predict a single intermediate feature $\mathcal{F}_\text{inter}^t \in \mathbb{R}^{768}$.
   \item \emph{Token Generation}: This feature $\mathcal{F}_\text{inter}^t$ conditions the flow matching network $\mathcal{F}_\text{flow}$. $\mathcal{F}_\text{flow}$ generates the next $\mathbb{R}^{32}$ B-rep latent token $H_a^t$ by progressively denoising a random Gaussian sample over 100 internal timesteps.
   \item \emph{Iteration}: The newly generated $H_a^t$ is projected to $\mathbb{R}^{768}$, appended to the input history, and the process repeats for step $t+1$. This continues until the transformer predicts a special end-of-sequence (EOS) token.
\end{itemize}

\section{Results and Evaluation}

\subsection{Dataset Evaluation}
\label{subsec:dataset_eval}
To the best of our knowledge, CAD-Editor~\cite{yuan2025cadeditor} is the only other existing dataset targeting text-driven B-rep editing. We therefore conducted a user study to compare our \dname with CAD-Editor. 
We randomly sampled 30 editing cases from each dataset and asked 20 participants to evaluate the quality of the editing scenarios. 
Participants were shown pairs of scenarios (one from each dataset), with each scenario consisting of the pre-edit model, the post-edit model, and the corresponding text prompt. 
They were then asked to choose the scenario they preferred based on four criteria:
(i)~\textbf{Practical utility}: which operation represents a more realistic task likely to be used in common CAD workflows.
(ii)~\textbf{Prompt alignment}: which text prompt more accurately and completely describes the visible changes.
(iii)~\textbf{Edit complexity}: which editing task appears to be more difficult or involved in performing.
(iv)~\textbf{Model complexity}: which shape has more intricate structural details. 
The results are summarized in \Cref{fig:res_userstudy}. Overall, our \dname dataset was strongly favored over CAD-Editor across all aspects. 
CAD-Editor dataset is constructed using sketch-and-extrude operations and usually contains only a single extrusion, making the models and edits relatively simple. In contrast, our \dname is built directly upon B-rep models and is carefully prompted to include rich functional descriptions, resulting in more diverse and complex B-rep models and edits, see \Cref{fig:res_dataset}.

\subsection{Method Evaluation}
\label{subsec:comp}
\paragraph{Baselines.} There is no existing method that directly targets text-driven CAD editing without its construction history. Therefore, to establish a comparison, we evaluate our method against the state-of-the-art approach, CAD-Editor~\cite{yuan2025cadeditor}, on their released CAD-Editor dataset. While this dataset provides construction history (which CAD-Editor requires), our method does not use it. \rev{We also performed preliminary training on the CAD-Editor dataset, but found that the smaller scale of that dataset causes our high-capacity autoregressive model to overfit quickly.}

We also include several unconditional generative models (HNC-CAD~\cite{hnccad}, BrepGen~\cite{brepgen24}, HoLa-BRep~\cite{HolaBRep25}). As these methods cannot perform editing tasks, we only report their realism metrics for context.

\rev{
\paragraph{Metrics.}
Given a pre-edit B-rep and text prompt from the CAD-Editor testset, we evaluate CAD-Editor and our method under the same protocol. We follow the same metrics used in CAD-Editor, namely Validity, D-CLIP, JSD, and CD, and additionally report human preference. For each input, both editing methods generate 16 edited B-rep candidates. We mark an input case as valid if at least one of the 16 candidates is a valid B-rep, and report the percentage of such cases as \textit{Validity}. For D-CLIP, CD, and the human study, we randomly select one valid candidate from the 16 whenever available. 
For the unconditional generative baselines, which do not perform text-driven editing, we only report the realism metrics JSD and CD for context.
\begin{itemize}
    \item D-CLIP score~\cite{yuan2025cadeditor}: Measures whether the directional change between the source and target model latents aligns with the input prompt's embedding.
    \item Jensen-Shannon Divergence (JSD)~\cite{deepcad21,yuan2025cadeditor}: Measures the realism of the generated models compared to the real distribution.
    \item Chamfer Distance (CD): Following HoLa-BRep and BrepGen, we use CD as a distribution-similarity metric by reporting the minimum CD between 3000 generated models and 1000 ground-truth references.
    \item Human: The percentage of time users prefer the generated results over CAD-Editor.
\end{itemize}
}

\begin{figure}[t]
\centering
\includegraphics[width=\linewidth]{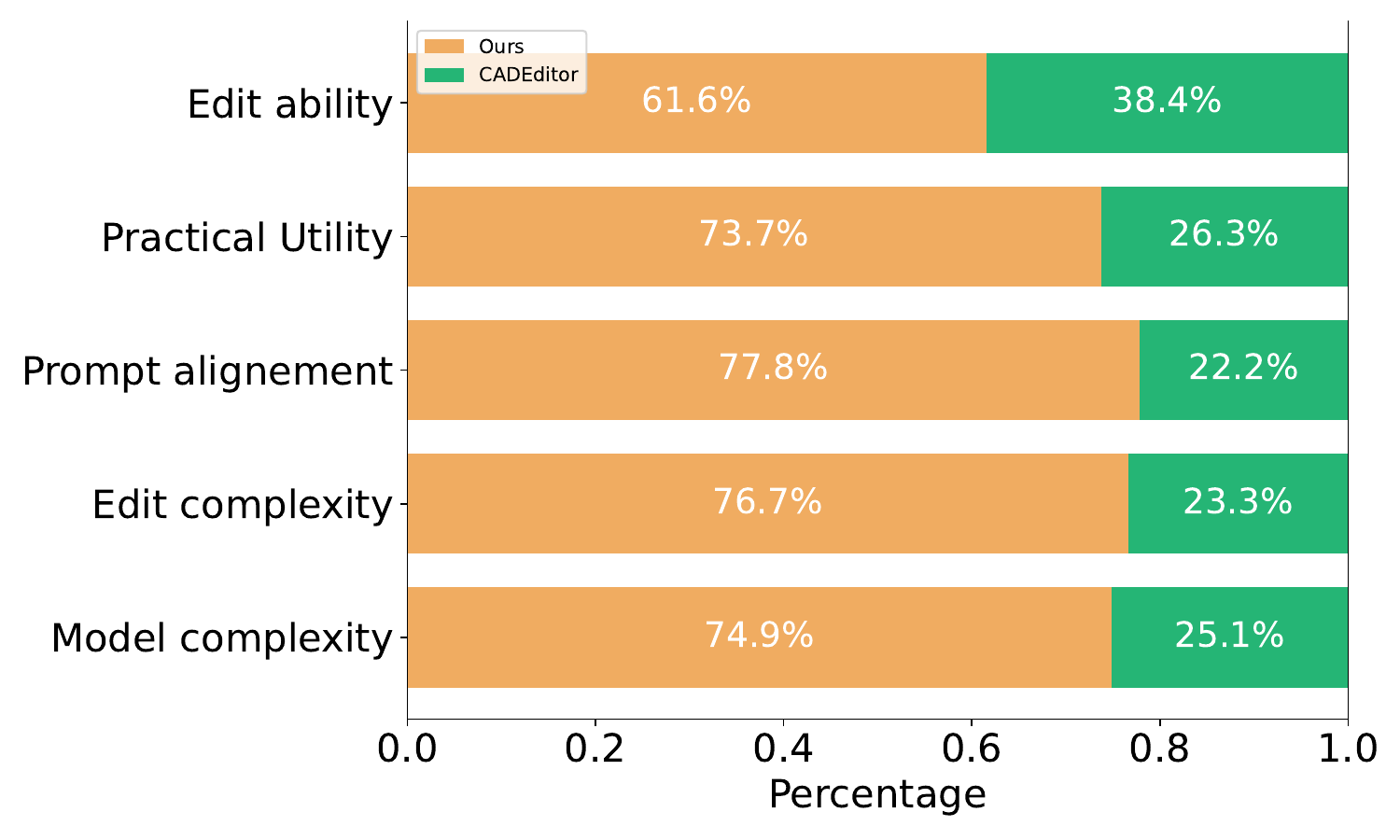}
\caption{\textbf{User study.} We ask 20 participants to compare our \dname dataset in terms of practical utility, prompt alignment, edit complexity, and model complexity; and our algorithm in terms of edit ability -- we compare against CAD-Editor~\cite{yuan2025cadeditor}. Our dataset and algorithm are both preferred across all aspects.
}
\label{fig:res_userstudy}
\end{figure}

\begin{table}[t]
    \caption{Comparison with existing methods on CAD-Editor testset.}
    \label{tab:comparison}
    \centering
    \resizebox{\columnwidth}{!}{
        \begin{tabular}{r|c|c|c|c|c}
            \toprule[0.4mm]
            \multirow{2}{*}{Method}       & \multicolumn{2}{c|}{Realism}   & \multicolumn{2}{c|}{Editability} & \multirow{2}{*}{Human} \\
                   & JSD ↓   & CD ↓ & Validity ↑ & D-CLIP ↑ \\ \midrule
            HNC-CAD      & 2.11   & 2.25 & - & - & - \\
            BrepGen      & 2.20   & 1.70 & - & - & - \\
            HoLa-BRep    & 1.66   & 1.99 & - & - & - \\
            \midrule
            CAD-Editor    & 3.06   & 2.11 & \textbf{99.79\%} & 0.25 & 38.4\% \\
            Ours         & \textbf{2.96}   & \textbf{1.76} & 97.13\% & \textbf{0.34} & 61.6\% \\

            \bottomrule[0.4mm]
        \end{tabular}
    }
\end{table}

For the user study, we \textit{randomly} sample 10 editing cases from the CAD-Editor testset and ask 20 participants to evaluate which edited result better follows the input text prompt while preserving the original model's identity.

As shown in \Cref{fig:res_userstudy}, and \Cref{tab:comparison}, our method outperforms CAD-Editor in terms of D-CLIP score and user preference, while maintaining a high validity ratio.

\begin{figure}[t]
\centering
\includegraphics[width=0.99\linewidth]{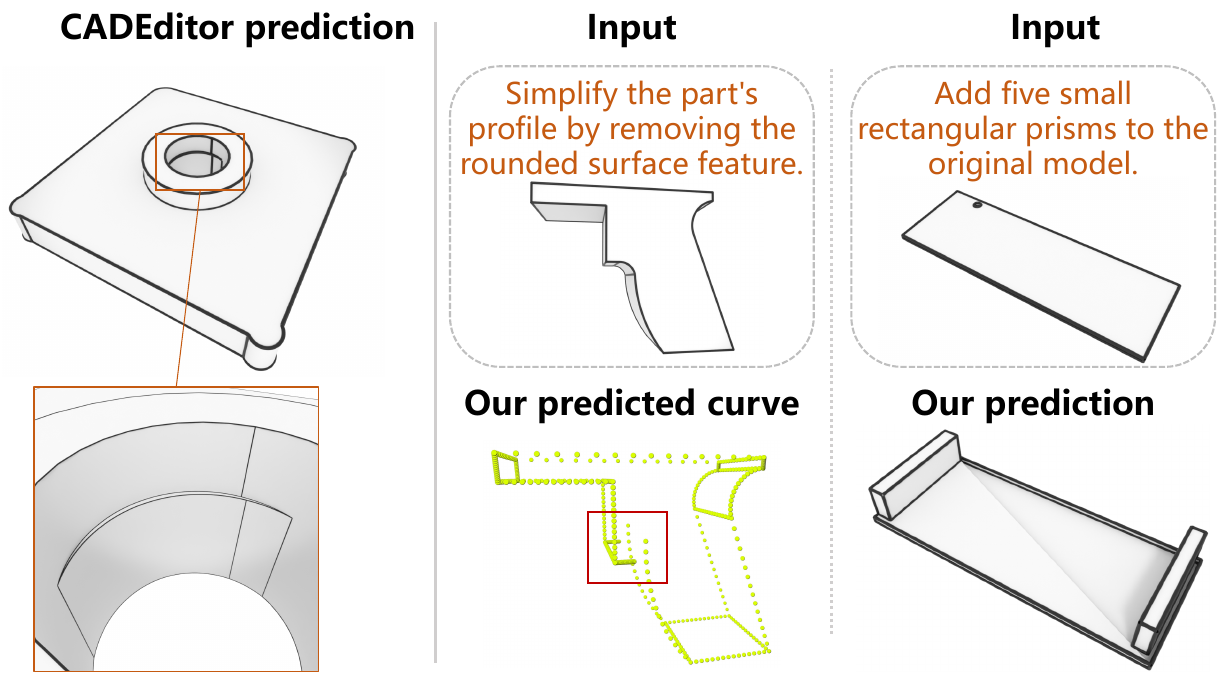}
\caption{\textbf{Failure cases.} \rev{Although CAD-Editor can generate valid and watertight CAD models, its multi-stage construction pipeline is sensitive to quantization errors. For example, the split cylinder (\textbf{Left}) is caused by misalignment in the quantized cylinder parameters across different stages, which leads to an undesired separation artifact. \name (\textbf{Right}) also suffers from invalid B-rep prediction and fails to interpret user intent when the editing involves counting.}}
\label{fig:failure_case}
\end{figure}

\begin{figure*}[t]
    \centering
    \includegraphics[width=0.9\textwidth]{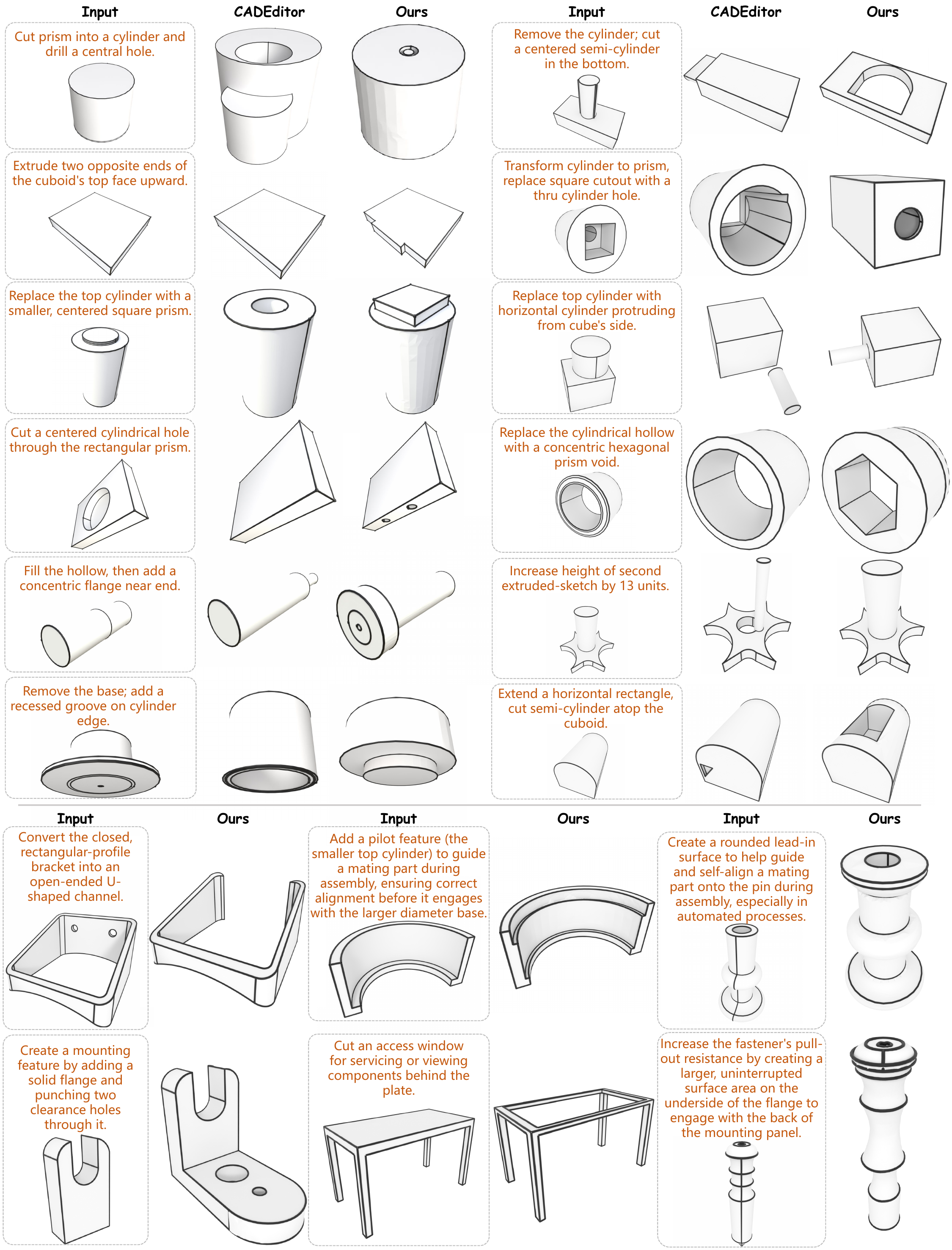}
    \caption{\textbf{Top left: } \textit{Randomly} selected comparison samples from CAD-Editor testset. \textbf{Top right: } Challenging comparison samples from CAD-Editor testset. \textbf{Bottom: } Challenging samples from our \dname testset. More results can be found in the supplementary.
    }
\label{fig:res_algorithm}
\end{figure*}

Qualitative comparisons are shown in \Cref{fig:res_algorithm} (top), where we show 5 random cases from the user study on the left and 5 additional challenging cases on the right. The sketch-and-extrude representation used in CAD-Editor not only limits its scalability to complex geometries (\eg B-spline primitives) and datasets lacking construction history (\eg ABC~\cite{ABC}), but it also introduces other problems. The quantization required for tokenizing CAD primitives, especially coordinates, can lead to large precision loss and significant deviations from the intended design. As shown in \Cref{fig:failure_case}, CAD-Editor often generates misaligned components due to quantization error and unstable token prediction. Some edits even result in two separate models (\Cref{fig:res_algorithm}, third row, right).

In contrast, ours demonstrates an improved capability to capture fine-grained geometric details. This leads to more accurate and reliable editing results on more complex shapes and even high-level and functional editing instructions, as shown in the bottom of \Cref{fig:res_algorithm}. \rev{Notably, although our training pairs are synthesized from delete operations, the model still generalizes to semantically richer instructions such as replacing, enlarging, or increasing a part.}

\subsection{Ablation}
\label{subsec:abl}
To validate our variational autoregressive transformer, we conduct an ablation study comparing it against two alternatives: (i) a deterministic autoregressive transformer and (ii) a pure flow matching network (similar to BrepGen~\cite{brepgen24} and HoLa-BRep~\cite{HolaBRep25}). We experimented on a subset of the \dname testset, evaluating the generated B-rep models using the same metrics as in Sec.~\ref{subsec:comp}. \rev{Here we use 8-choose-1 protocol for efficiency.}
The results are in \Cref{tab:ablation}. Our full method achieves the best performance in both validity and D-CLIP score.
When replacing our variational transformer with a deterministic one, the validity drops significantly. This is because the intrinsic ambiguity in text-driven editing requires the model to generate diverse, plausible outputs, a task ill-suited for a deterministic model.
On the other hand, the pure flow matching network struggles because the B-rep latent space is variable-length. The necessary padding and unpadding operations introduce noise during training and inference, leading to inferior performance in both metrics.
Our full variational autoregressive transformer effectively handles the variable-length B-rep latent features while simultaneously capturing the inherent uncertainty in text-driven editing, resulting in the best performance.

\begin{table}[t]
    \caption{Ablation on the variational autoregressive transformer.}
    \label{tab:ablation}
    \centering
    \renewcommand{\arraystretch}{1.2}
        \begin{tabular}{r|c|c}
            \toprule[0.4mm]
            Method   & Validity ↑ & D-CLIP ↑ \\ \midrule
            Full     & \textbf{78.4\%}  &  \textbf{0.91}       \\
            w/ deterministic autoregressive  & 45.6\%   & 0.50                   \\
            w/ flow matching & 65.8\%                    & -0.94                   \\
            \bottomrule[0.4mm]
        \end{tabular}
\end{table}

\subsection{Limitation and Discussion}
\label{subsec:discuss}
\rev{Our approach, by modeling the latent space directly, can sometimes generate misaligned surface patches or curves, resulting in an invalid B-rep. The editing performance is still constrained by the underlying HoLa-BRep latent space and decoder. In some failure cases, the predicted latent features reflect the intended semantic change, but the final decoded UV grids still fail to form a watertight B-rep. At the same time, it is encouraging that this latent space, which is pre-trained purely on geometric signals without text supervision, still aligns well enough with language to support semantic editing.

In addition, while our data collection pipeline can capture complex geometric changes involving multiple B-rep faces, it does not explicitly model or enforce high-level geometric constraints. This contrasts with real-world CAD modeling, where design intent often relies on relationships like symmetry, orthogonality, or parallelism between parts. Because these constraints are not explicitly captured in our dataset, \name may fail to preserve them during an edit, potentially limiting its generalization in complex design scenarios.
}

\begin{figure*}[t]
    \centering
    \includegraphics[width=0.99\textwidth]{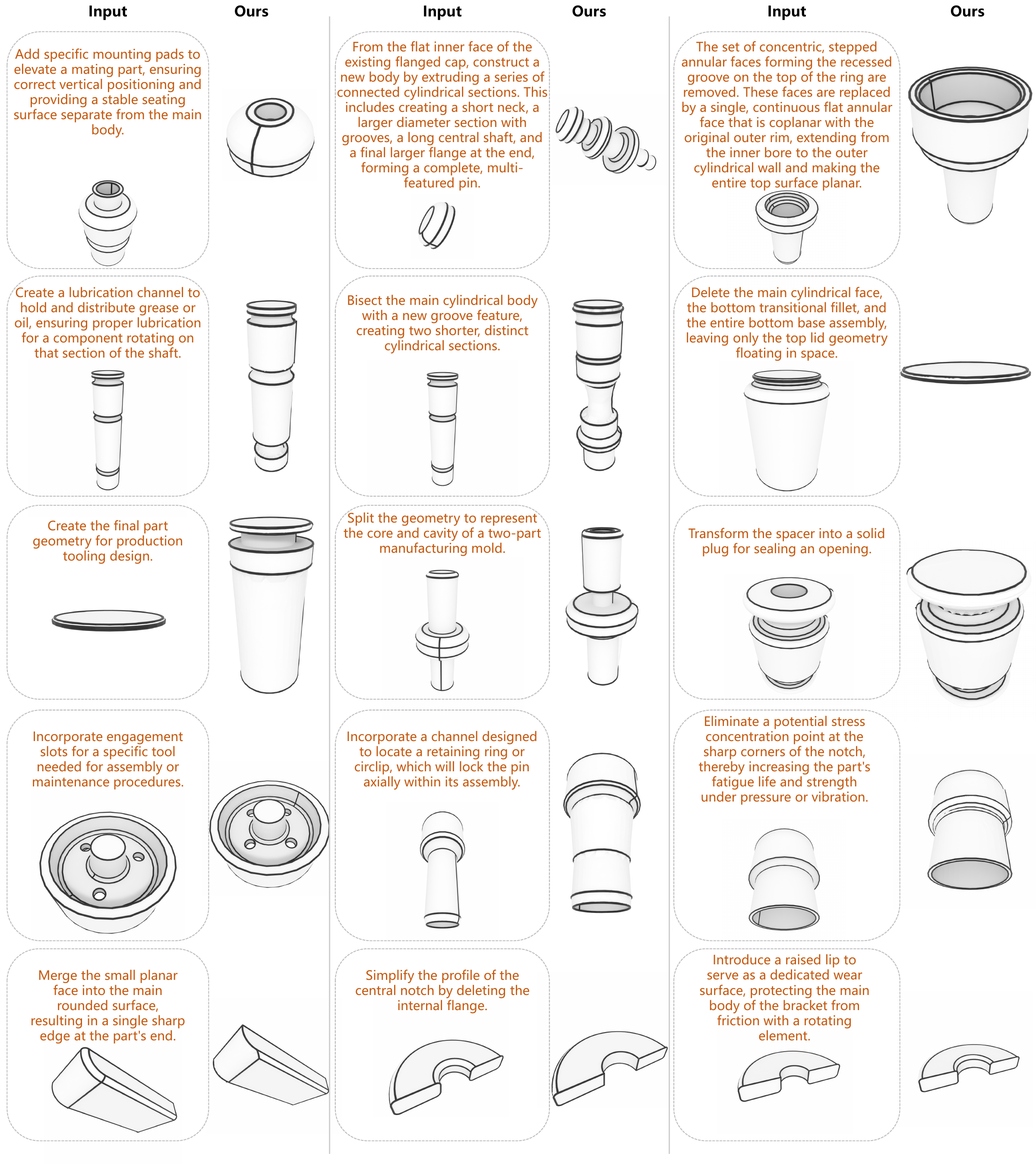}
    \caption{More examples of \name from our \dname dataset.
    }
\label{fig:results_more1}
\end{figure*}

\section{Conclusion}
\label{sec:conclusion}

We presented \name, the first method for direct text-driven editing of complex B-rep CAD models, including freeform surfaces, without requiring construction history. By editing in a learned B-rep latent space, our approach avoids the brittleness and expressiveness limitations of token-based pipelines while supporting variable-length geometry and one-to-many edits. We also introduced \dname, the first large-scale dataset for this task, constructed automatically with CAD operations and mLLM-based semantic annotation.

Experiments show that \dname contains richer and more practical edits than prior data, and that \name outperforms existing methods while using less input information. Meanwhile, challenges remain in counting-heavy prompts, complex spatial reasoning, and preserving high-level geometric constraints such as symmetry or parallelism. Future work includes stronger constraint-aware and physics-aware editing. \rev{On the data side, our pipeline can be extended through Fusion360 APIs to support richer edits, such as parameter changes and stacked operations. As CAD kernels already handle much of the low-level validity, a key next step is better semantic filtering to curate more practical and diverse edit pairs.}

\section*{Acknowledgments}
We thank the anonymous reviewers for their insightful comments, and Remy Sabathier and Rishabh Kabra for their valuable feedback and helpful discussions. We are also grateful to all the user study participants for their time and contributions. CL was supported by a gift from Adobe.

\bibliographystyle{ACM-Reference-Format}
\bibliography{main.bib}

\clearpage

\setcounter{section}{0}
\setcounter{table}{0}
\setcounter{figure}{0}
\makesupplementtitle

In the supplementary material, we provide further details regarding our variational autoregressive model (\Cref{sec:var_model}) and specific training parameters (\Cref{sec:training_details}). We also present additional dataset statistics (\Cref{sec:dataset_stats}) and the full list of prompts used in our experiments (\Cref{sec:prompts}). Finally, we showcase additional qualitative results.

\section{Variational Autoregressive Model}
\label{sec:var_model}
The task of language-guided B-rep editing is inherently \emph{one-to-many}: a single textual prompt can yield multiple diverse, yet valid, geometric outputs.
Furthermore, the B-rep latent sequences, as defined by HoLa-BRep~\cite{HolaBRep25}, are of \emph{variable length}.
To effectively model the distribution of possible edits while accommodating this variable-length structure, we designed a variational autoregressive model.

Our architecture, illustrated in Figure 5 of the main paper, combines a generative autoregressive transformer with a flow matching network.

\rev{
\paragraph{Multimodal Encoder.}
As detailed in the main paper (Section 4), the model is conditioned on a set of multimodal inputs: the source B-rep latent sequence $\{H_{b}^{i}\}$, the text instruction $T_{b\rightarrow a}$, a rendered image $I_{b}$, and a 2D bounding box $B$. These inputs are processed by their respective encoders (\eg DINOv2 for images, mLLM for text) and projected into a shared $768$-dimensional feature space. In the implementation, each modality-specific projection is a two-layer MLP with a ReLU nonlinearity between the layers. Concretely, the global image feature is mapped by $1024 \rightarrow 768 \rightarrow 768$, the text feature by $2560 \rightarrow 768 \rightarrow 768$, the per-face image feature by $1024 \rightarrow 768 \rightarrow 768$, the bounding box feature by $4 \rightarrow 768 \rightarrow 768$, and the source B-rep latent token by $32 \rightarrow 768 \rightarrow 768$.

The conditioning sequence contains three global tokens for the image, text, and bounding box, together with per-face tokens. The three global condition tokens are implemented as learned parameters of shape $1 \times 768$, one for each modality. Each per-face token is formed by summing the projected pre-edit B-rep latent feature with the aligned per-face visual feature extracted by RoIAlign from the image feature map. We then add the learned global tokens to the corresponding projected image, text, and bounding-box features, mask out zero-padded faces, and optionally apply learned absolute positional encodings with embedding dimension $768$ and maximum sequence length $200$ to both the encoder and decoder streams.

The conditioning sequence is passed through a Transformer encoder implemented using PyTorch's \texttt{TransformerEncoder}. Specifically, the encoder contains $8$ stacked \texttt{TransformerEncoderLayer} blocks with model dimension $768$, $8$ attention heads, feed-forward dimension $2048$ and dropout rate $0.1$. A final \texttt{LayerNorm} is applied to the encoder output. The resulting context $\mathcal{F}_{src}$ is injected into the decoder via cross-attention.

During training, we further regularize the conditioning stream with modality dropout: the global image feature, the per-face visual features, and the optional bounding-box feature are randomly zeroed, and for a subset of samples the full encoder memory is masked to encourage robustness to missing conditions.

\begin{figure}[t]
    \centering
    \includegraphics[width=0.99\linewidth]{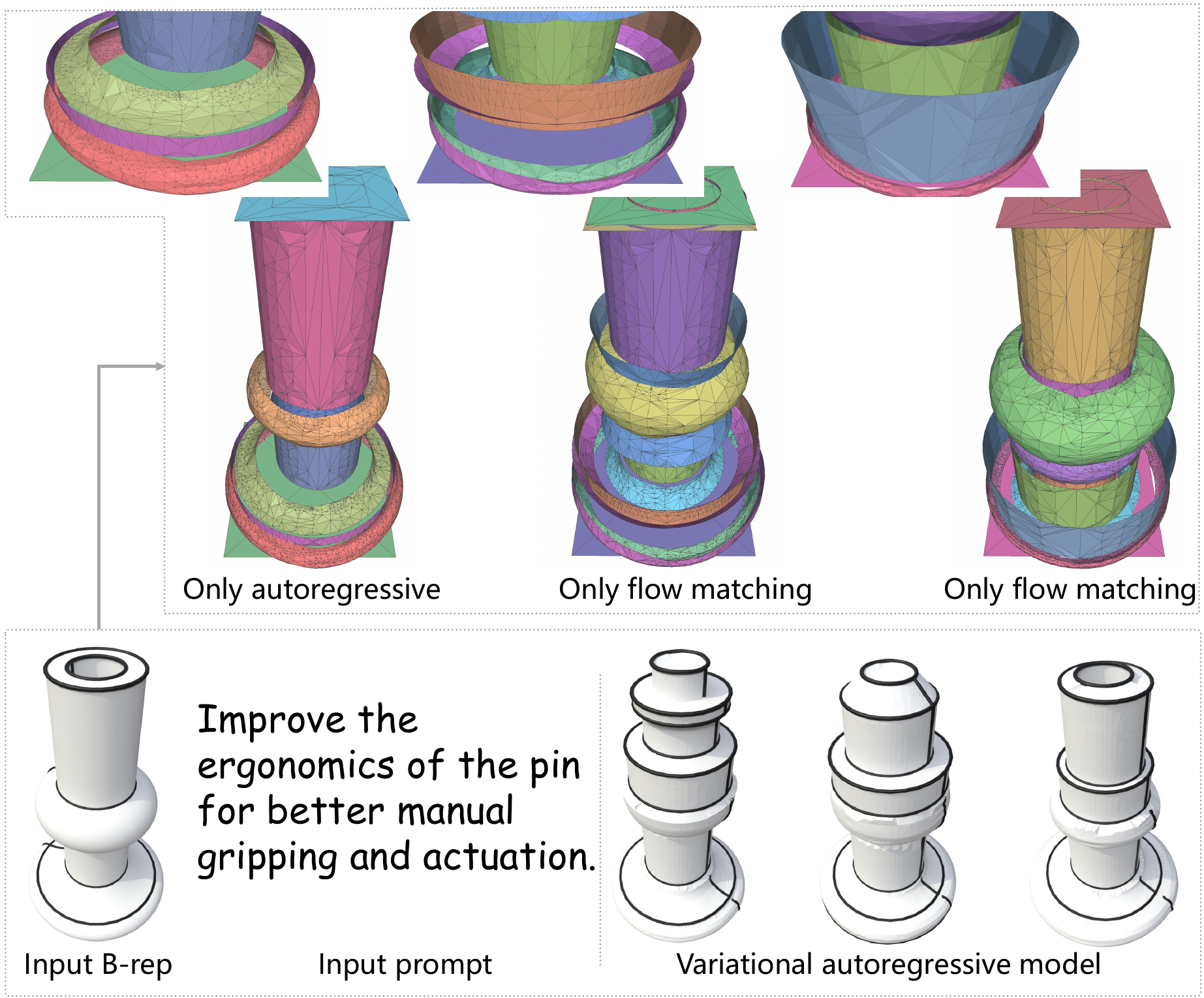}
    \caption{\textbf{Comparison of ablations.}
        Input B-rep model and editing instruction are shown in \textbf{bottom left}.
        Results from a vanilla autoregressive transformer failed to model the one-to-many distribution and usually produce results with missing geometry (\textbf{top left}).
        While a flow matching model can produce multiple outputs, the padding and unpadding operations cause issues and produce either duplicated or missing surfaces, resulting in invalid B-reps (\textbf{top right}).
        Our variational autoregressive model produces diverse and valid edits that align well with the instruction (\textbf{bottom right}).
    }
    \label{fig:ablation_comparison}
\end{figure}

\paragraph{Variational Decoder and Latent Sampling.}
The decoder is implemented using PyTorch's \texttt{TransformerDecoder}. The input target token at each step is obtained by another two-layer MLP with ReLU, mapping $32 \rightarrow 768 \rightarrow 768$. Decoding starts from a learned start token of dimension $768$. The decoder contains $8$ stacked decoder layers with model dimension $768$, $8$ attention heads, feed-forward dimension $2048$ and dropout rate $0.1$, followed by a final \texttt{LayerNorm}. Each decoder block applies masked causal self-attention over the generated target sequence, cross-attention to the encoder memory $\mathcal{F}_{src}$, and a position-wise feed-forward network. During training, the decoder operates autoregressively from the learned start token together with the teacher-forced target face embeddings, and produces one decoder state per output position.

A critical aspect of our design is that the decoder does \emph{not} directly predict the discrete latent tokens. Instead, at each autoregressive step $t$, the decoder attends to the encoder context $\mathcal{F}_{src}$ and the previously generated tokens $\{H_{a}^{0},...,H_{a}^{t-1}\}$ to predict a single \emph{intermediate conditioning feature} $\mathcal{F}_{inter}^{t}$.

\begin{table}[t]
    \caption{\rev{Ablation on the component of the variational autoregressive transformer.}}
    \label{tab:ablation2}
    \centering
    \renewcommand{\arraystretch}{1.2}
    \begin{tabular}{r|c|c}
        \toprule[0.4mm]
        Method                                          & Validity ↑      & CD ↓          \\ \midrule
        Full                                            & \textbf{87.6\%} & \textbf{0.061} \\
        w/o image feature $\mathcal{F}_{\text{img}}$    & 70.9\%          & 0.069          \\
        w/o bbox feature $\mathcal{F}_{\text{bbox}}$    & 85.4\%          & 0.064         \\
        w/o face roi feature $\mathcal{F}_{\text{roi}}$ & 84.1\%          & 0.435         \\
        \bottomrule[0.4mm]
    \end{tabular}
\end{table}

\begin{figure*}[t]
    \centering
    \includegraphics[width=1\linewidth]{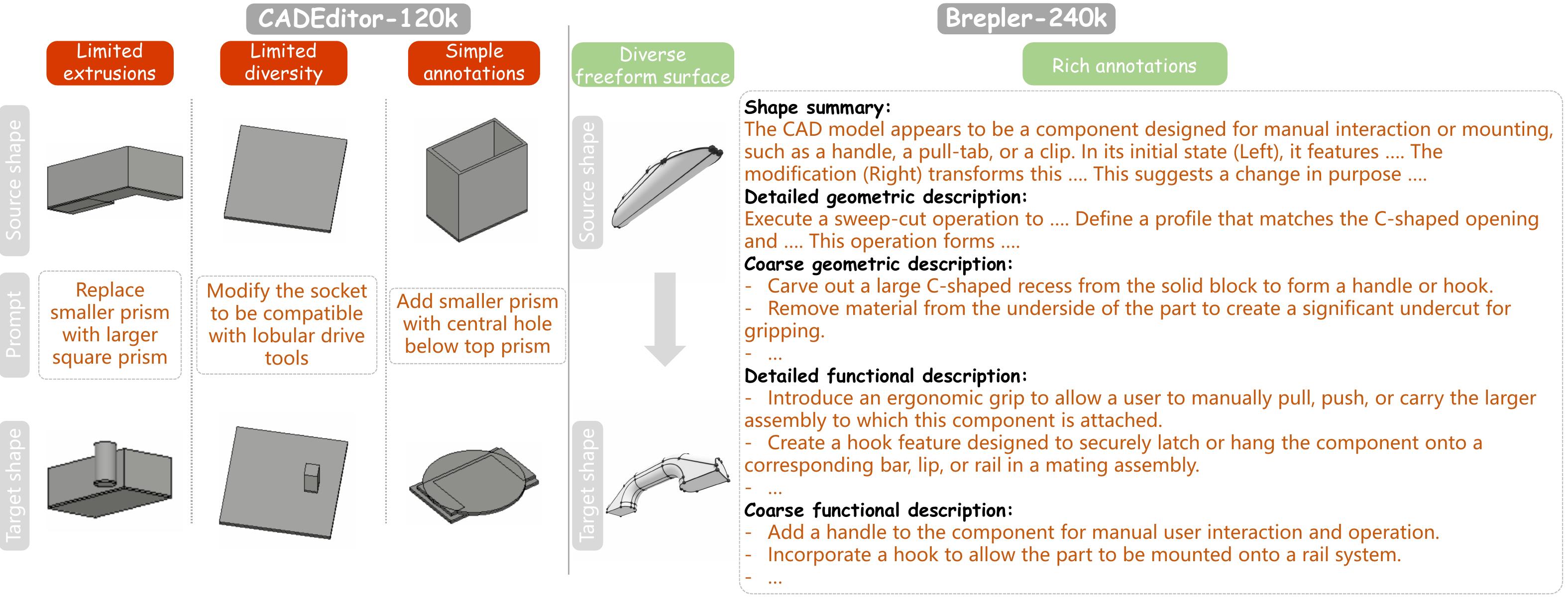}
    \caption{\textbf{Dataset motivation.} Existing datasets, such as CAD-Editor, are often limited by the sketch-and-extrude pipeline and typically contain only single-extrusion data sequences, resulting in limited data variety. In contrast, \dname directly models the B-rep, thus supporting complex geometry, and providing rich, multi-level annotations.}
    \label{fig:dataset_motivation}
\end{figure*}

\begin{figure*}[t]
    \centering
    \includegraphics[width=0.99\textwidth]{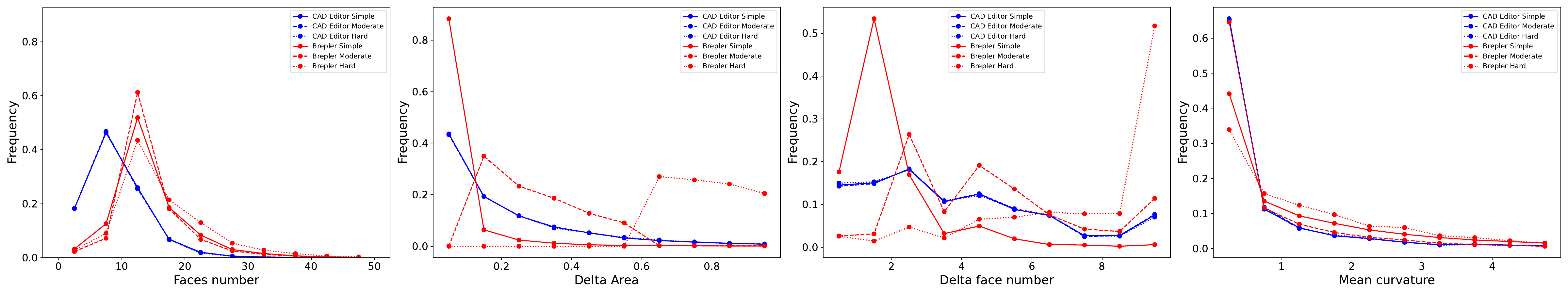}
    \caption{\textbf{Dataset statistics.} Statistics of the \dname dataset and split it into three subsets based on the complexity of the edits: \emph{Simple}, \emph{Moderate}, and \emph{Hard}. We also use the same criteria to split the CAD-Editor dataset for comparison.
    }
    \label{fig:dataset_stats}
\end{figure*}

This intermediate feature $\mathcal{F}_{inter}^{t}$ conditions a small flow matching network, which aims to generate the next B-rep latent token $H_{a}^{t} \in \mathbb{R}^{32}$. It does so by learning a continuous flow from a simple Gaussian noise sample $z \sim \mathcal{N}(0, I)$ to the target token $H_{a}^{t}$, conditioned on $\mathcal{F}_{inter}^{t}$ and a time step embedding. This variational approach effectively models the rich distribution of possible next tokens, addressing the one-to-many nature of the editing task.

Concretely, during training we sample a time step from a Gaussian, map it to $[0,1]$ using a sigmoid transform, and quantize it into $1000$ discrete levels. This time step is used to interpolate between the target token $H_a^t$ and Gaussian noise. The interpolated noisy token is first projected by a linear layer $32 \rightarrow 768$. In parallel, the time step is embedded by a learned time-embedding module that outputs a $768$-dimensional feature. The noisy-token feature and the time embedding are then combined with the conditioning feature $\mathcal{F}_{inter}^{t}$ and processed by a flow matching network composed of $6$ DiT-style AdaLN blocks, each configured with hidden dimension $768$, conditioning dimension $768$, and $8$ attention heads in the implementation. A final linear layer maps the resulting feature back to $\mathbb{R}^{32}$ to predict the velocity field. The network is optimized using a standard L2 loss on the predicted vector field, as in standard conditional flow matching. Similar to MAR~\cite{autotrans24}, each latent token is generated independently by this network, conditioned on its corresponding feature from the transformer decoder.

\paragraph{EOS Classification.}
To handle variable-length outputs, we train an additional binary classifier implemented as a linear layer $768 \rightarrow 1$ on the decoder's output features. This classifier is optimized with a binary cross-entropy (BCE) loss to predict whether the current step $t$ should be the \emph{End-of-Sequence (EOS)}. During inference, the decoder generates autoregressively until this classifier predicts the EOS token with high confidence ($0.5$), and each new face token is obtained by numerically integrating the learned flow from an initial Gaussian sample over $100$ discrete denoising steps.
}

\paragraph{Discussion.}
As shown in \Cref{fig:ablation_comparison}, our variational autoregressive design effectively captures the diversity of possible edits while maintaining high fidelity to the input prompt and source geometry.

In contrast, alternative architectures struggle with the task's inherent complexities. A vanilla autoregressive transformer, lacking the flow matching component, fails to model the complex, one-to-many output distribution, frequently leading to invalid outputs. A pure flow matching network is similarly ineffective, as its performance is compromised by the padding and unpadding operations required to handle variable-length sequences, which also degrades the validity.

\rev{
We also ablated the contribution of each input modality on 500 randomly sampled test examples from our \dname dataset, and computed the validity and Chamfer Distance (CD) between the generated and ground-truth B-rep models. As shown in \Cref{tab:ablation2}, removing any of the input features leads to a drop in performance.
Notably, the training of the variational autoregressive model becomes unstable without the face ROI feature $\mathcal{F}_{\text{roi}}$, which provides crucial local image context to the B-rep latent features. This highlights the importance of this feature in bridging the gap between the 2D image and 3D geometry, and its role in stabilizing the training process.
}

\section{Training Details}
\label{sec:training_details}
We trained the Transformer from scratch using the AdamW optimizer with a batch size of 192 and a learning rate of $1\times10^{-4}$. The model was trained on 8 NVIDIA RTX 5090 GPUs with half precision, and training converged in approximately 48 hours.

To improve model robustness and prevent overfitting to any single modality, we applied a regularization strategy during training: the rendered image feature ($\mathcal{F}_{img}$), the per-face RoI feature ($\mathcal{F}_{roi}$), and the optional bounding box feature ($\mathcal{F}_{bbox}$) were randomly masked (set to a zero vector) with a probability of $0.5$. In addition, for a subset of training samples, we masked the full encoder memory.

For the flow matching network, we used discrete time step sampling during training by drawing a sigmoid-transformed Gaussian time step and quantizing it into $1000$ levels. At inference, each B-rep latent token $H_{a}^{t}$ is generated by solving the flow ODE over $100$ discrete denoising steps.

The model has 167M trainable parameters. The generation process takes approximately 2 seconds per shape on an NVIDIA RTX 5090 GPU without particular optimization like KV-cache.

\section{Dataset Statistics}
\label{sec:dataset_stats}

\begin{figure*}[t]
    \centering
    \includegraphics[width=0.99\textwidth]{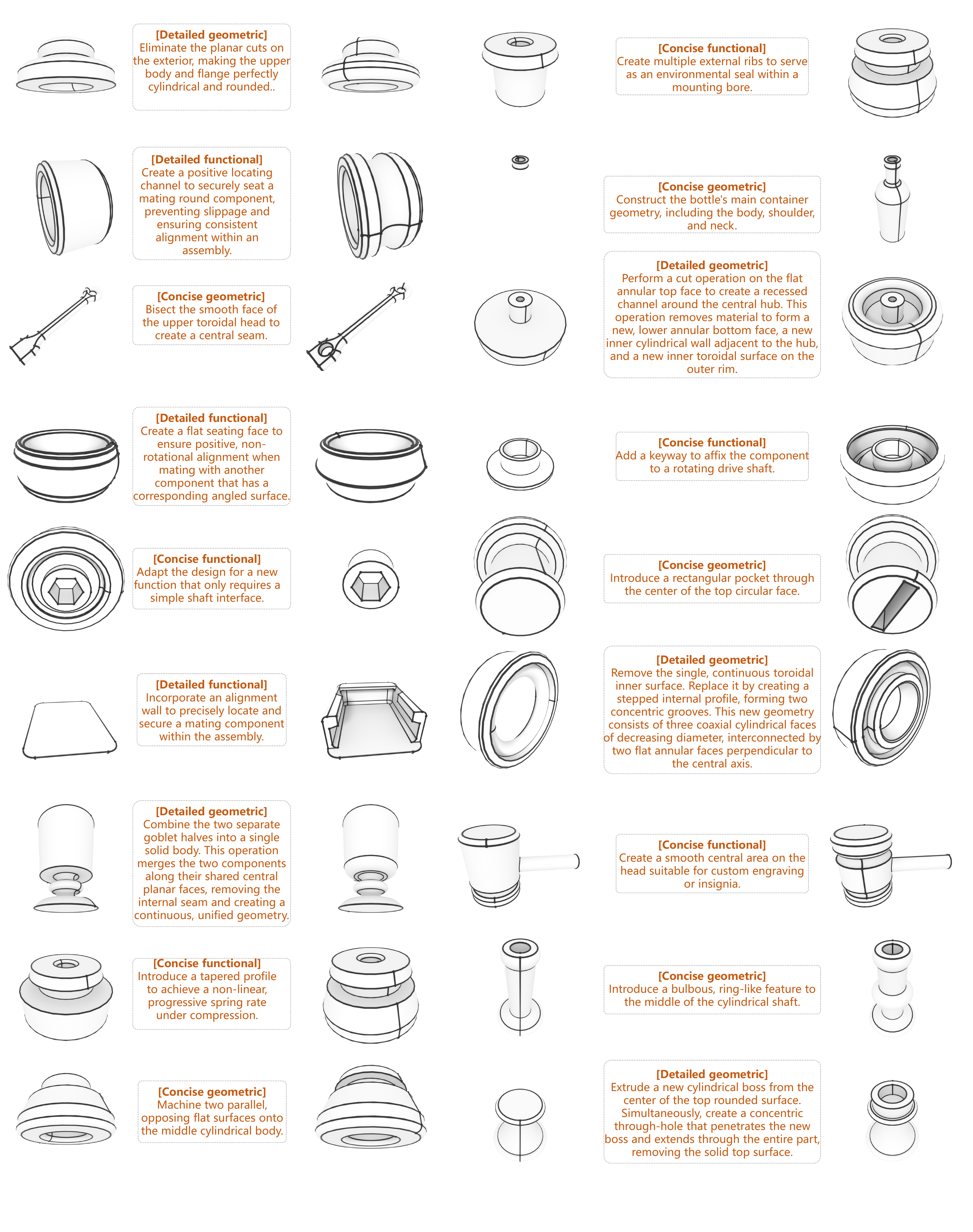}
    \caption{More samples from \dname dataset.
    }
    \label{fig:dataset_samples1}
\end{figure*}

\begin{figure*}[t]
    \centering
    \includegraphics[width=0.99\textwidth]{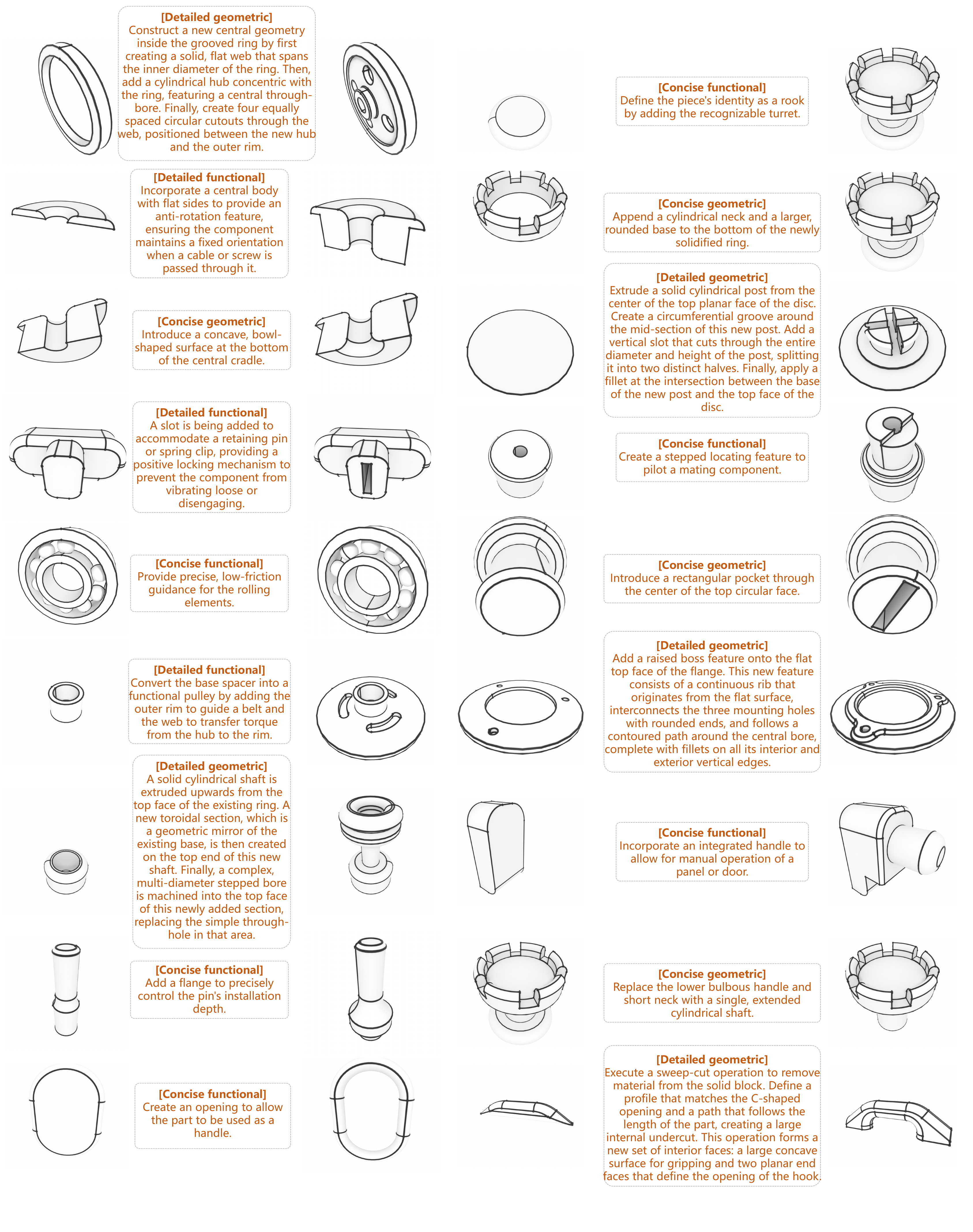}
    \caption{More samples from \dname dataset.
    }
    \label{fig:dataset_samples2}
\end{figure*}

We show additional statistics of the \dname dataset in \Cref{fig:dataset_stats}, including distributions of the number of surfaces of the source B-rep models, the area ratio of the edited region, the number of changed surfaces after editing, and the mean curvature of the edited surfaces.
Our dataset covers a wide range of geometric complexities (both prismatic and freeform) and editing scenarios, demonstrating its suitability for training and evaluating language-guided B-rep editing models. We also provide a dataset comparison with CAD-Editor~\cite{yuan2025cadeditor} in \Cref{fig:dataset_motivation} and more dataset samples in \Cref{fig:dataset_samples1,fig:dataset_samples2}.

\rev{Our data construction follows a synthesize-then-filter strategy. For each input B-rep, we first enumerate valid face-deletion results using the CAD kernel, which guarantees that each retained pair remains topologically valid after local healing. We then filter this candidate pool and keep the case with the largest overall geometric change, measured by the total area of altered faces. This preference for larger yet still valid edits biases the dataset toward more informative editing pairs. The resulting distributions in \Cref{fig:dataset_stats}, especially the changed-face count and edited-area ratio, reflect this filtering strategy and show that the retained edits are not dominated by trivial local perturbations.}

We further categorize the dataset into three subsets based on the complexity of the edits:
\begin{itemize}
    \item \textbf{Simple edits}: Edits that involve only 1 surface changed or the changed area ratio less than 10\%.
    \item \textbf{Moderate edits}: Edits that involve more than 1 surface changed with the changed area ratio less than 60\%.
    \item \textbf{Hard edits}: Edits that involve more than 1 surface changed with the changed area ratio greater than 60\%.
\end{itemize}
This categorization allows for a more nuanced evaluation of model performance across varying levels of editing difficulty. We also use the 2:2:1 (simple/moderate/hard) ratio to construct the testset for a balanced evaluation.

\rev{\paragraph{Additional User Study Details.}
Our main-paper user studies involve 20 participants in total, among whom 4 are CAD domain experts with 5--10+ years of experience. Restricting the analysis to these experts further increases the preference for our method and dataset: for algorithm comparison, the expert-only edit-ability preference rises to 67.5\% (compared with 61.6\% overall); for dataset comparison, the expert-only preference for our annotations increases across all dimensions, including practical utility (86.7\%), prompt alignment (87.5\%), edit complexity (85.8\%), and model complexity (89.1\%).

For the algorithm comparison in the main paper, participants reviewed 10 shuffled pairs sampled from the CAD-Editor testset. To reduce the confounding effect of model complexity, we sampled these examples using a 1:1:1 ratio of simple, moderate, and hard edits according to the complexity split in \Cref{fig:dataset_stats}. For the dataset comparison study, participants reviewed 30 pairs, one from each dataset per trial.

We also conducted an additional supplementary expert study to disentangle annotation quality from model complexity. In this study, 4 CAD experts compared the annotations using 10 identical model pairs from CAD-Editor, so that only the text quality differed. Across all 10 examples, the experts unanimously preferred our annotations, suggesting that the gain is not only due to different model complexity, but also due to improved annotation quality.}

\section{Annotation}
\label{sec:prompts}
We provide the full prompt used to generate the editing instructions in \Cref{prompt:cad-editing}. The design of this prompt was an iterative process aimed at addressing several key challenges inherent in using mLLMs for CAD editing tasks.

\paragraph{Ambiguous Operational Order.}
Initial experiments revealed that providing the mLLM with rendered images of both the pre-edit ($M_b$) and post-edit ($M_a$) states often led to confusion. The model would misinterpret which state was the source and which was the target, resulting in incorrect or reversed editing instructions.
We mitigated this by concatenating the two images ($I_b$ and $I_a$) side-by-side and explicitly overlaying the text ``Left" and ``Right" onto the respective images. The prompt then clearly instructs the mLLM to analyze the geometric difference from the ``Left" state to the ``Right" state (or vice-versa), thereby establishing an unambiguous operational order.

\paragraph{Information leakage.}
When both states are visible, the mLLM sometimes ``cheats" by directly referencing visual features from the target state when describing the action to be taken on the source state. This results in instructions that are not self-contained, as they rely on the (future) visual context of the target, which is unsuitable for a standalone editing command.
We explicitly instructed the mLLM in the prompt to generate descriptions that are ``self-contained" and avoid phrases that refer explicitly to the visual features of the other image. This ensures the generated text instruction is meaningful and actionable without needing to see the resulting edit.

\paragraph{Difficulty of Locating the Edited Region.}
Without clear guidance, mLLMs struggled to identify the specific region of the CAD model that had been modified, particularly when the geometric difference was subtle or located on a complex freeform surface.
To focus the model's attention, we provided the 2D bounding box ($B$) of the changed area as an additional input. The prompt instructs the model to concentrate its analysis on the geometric features within this bounding box, while also noting that the change may affect adjacent areas.

\paragraph{Discussion.}
We also note that the geometric reasoning capabilities of mLLMs have improved progressively. In early experiments with less advanced models, we observed significant difficulty in accurately identifying and describing geometric changes, and the models often failed to follow complex, multi-part prompts.
However, the mLLM used in our final data generation pipeline (Gemini 2.5 Pro) demonstrated a substantial improvement. It was able to follow our detailed, multi-step prompt and generate the high-quality, multi-level annotations required for our dataset. This includes not just coarse geometric descriptions but also detailed functional inferences.
We believe this demonstrates that modern mLLMs possess a sufficient level of geometric and semantic reasoning to be highly effective for tasks that require high-level functional understanding beyond pure geometry processing. We anticipate this trend will continue, enabling even more sophisticated applications in the future.

\begin{figure*}[t!]
    \centering
    \includegraphics[width=0.99\textwidth]{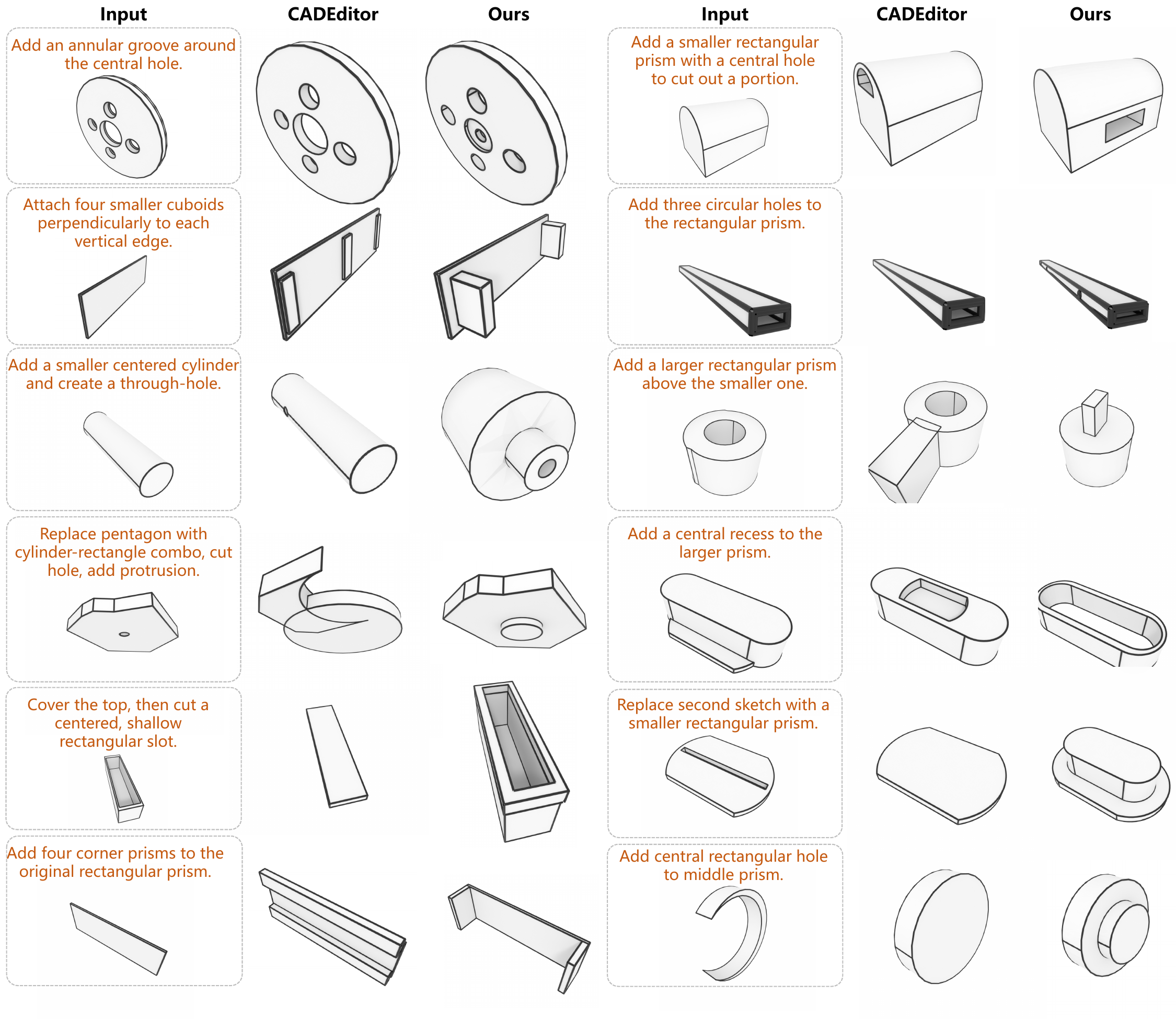}
    \caption{More comparison between \name and the baseline method on CAD-Editor testset.
    }
    \label{fig:qualitative_cadeditor}
\end{figure*}

\begin{figure*}[t]
    \centering
    \includegraphics[width=0.99\textwidth]{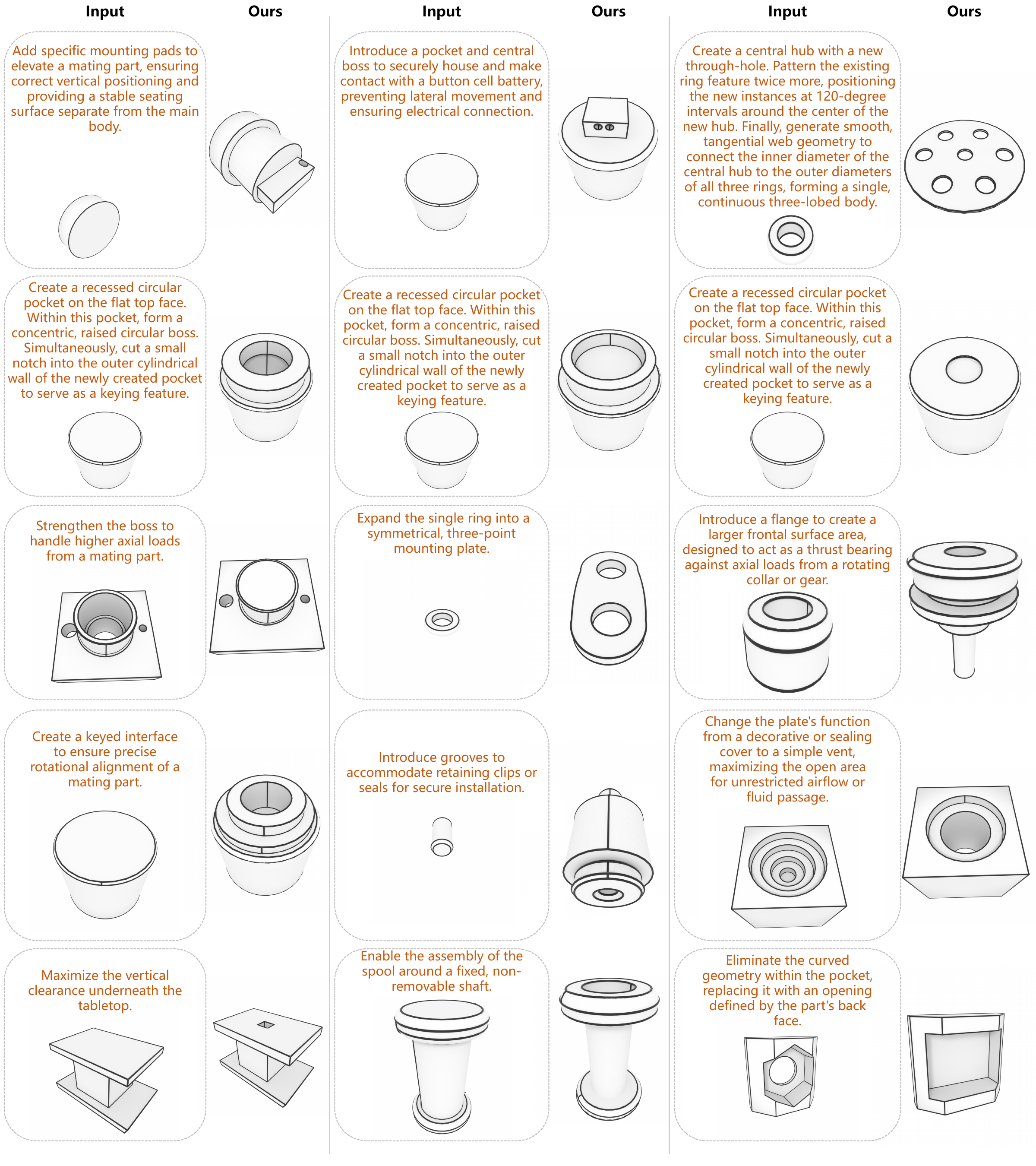}
    \caption{More generated editing results of \name on examples from our \dname dataset.
    }
    \label{fig:results_more2}
\end{figure*}

\begin{figure*}[t]
    \centering
    \includegraphics[width=0.99\textwidth]{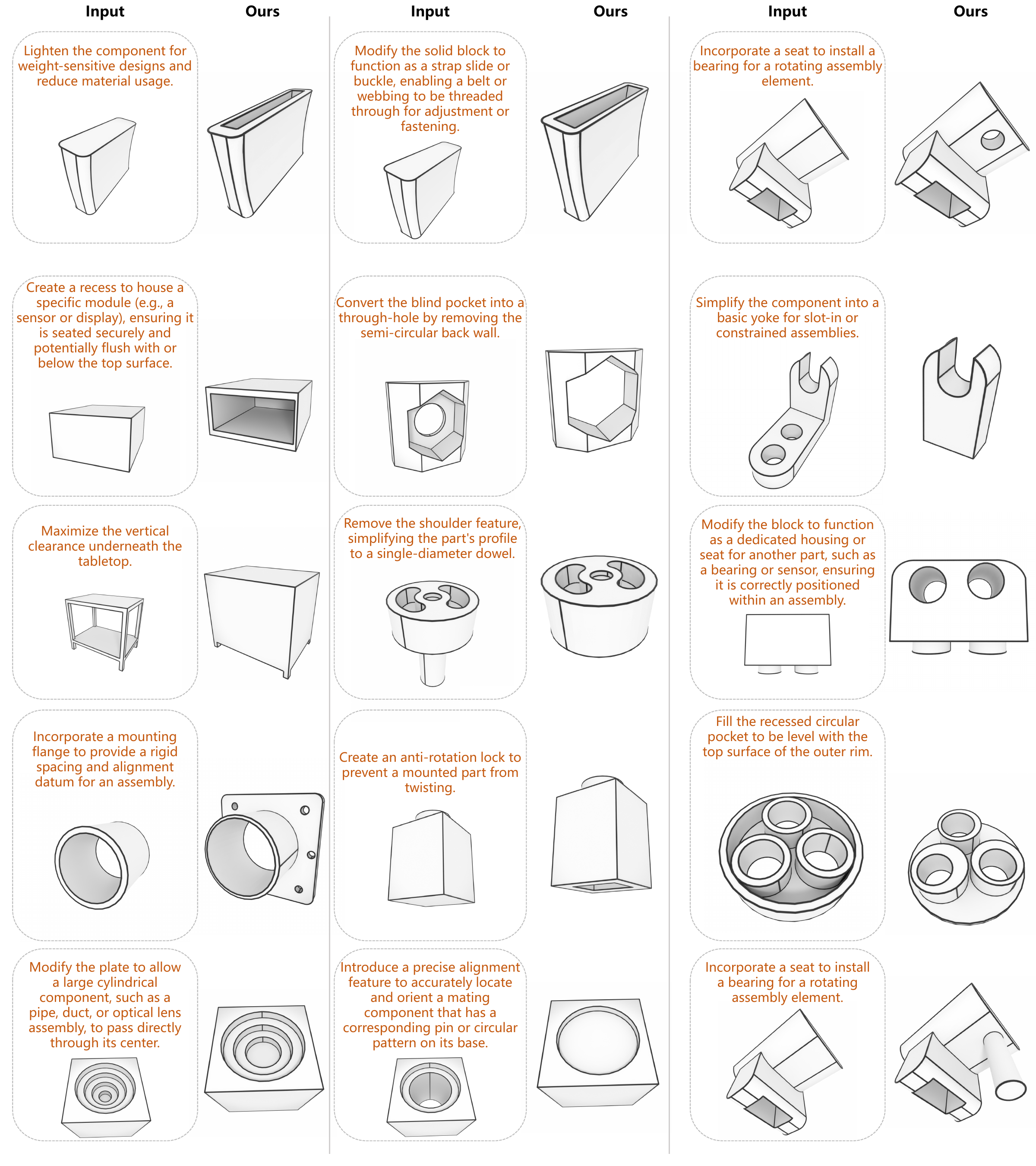}
    \caption{More generated editing results of \name on examples from our \dname dataset.
    }
    \label{fig:results_more3}
\end{figure*}

\begin{lstlisting}[caption={Example prompt to the mLLM}, label={prompt:cad-editing}]
**Prompt Objective:** Analyze a concatenated CAD image pair showing two states (Left and Right). Identify the geometric difference(s). Generate several descriptions based on the following requirements.
**Input Image Analysis (No output):**
Analyze the provided concatenated image.
* The **left half of the image** shows State A of a CAD model.
* The **right half of the image** shows State B of the same CAD model.
* The primary area where State A and State B differ is indicated by a **red bounding box/rectangle**. (Note: Use the box to focus your analysis, but understand it might be shifted and the difference could affect adjacent areas).
* Do not mention the bounding box in your output.

**Summary (Require output):**
* Output under a main heading: `## Summary:`
* Observe the images and infer the usage of the CAD model. Consider its function, intended application, and any relevant context that can be deduced from the geometry.
* Identify and Describe Geometric Differences in details for Both Directions (left to right and right to left) (Output This Next):**

**Geometric change description (Require output):**
* Output under a main heading: `## Geometric Change:`*
* Meticulously compare the geometry within and around the bounding box on the Left (State A) and Right (State B) images.
* Objectively note the geometric features present/absent or changed between State A and State B.
* Under this main heading, provide **TWO distinct, detailed descriptions about the change**:
  * One under a sub-heading `### Left -> Right Change:` describing the **ALL changed faces** to change State A (Left) into State B (Right).
  * One under a sub-heading `### Right -> Left Change:` describing the **ALL changed faces** to change State B (Right) into State A (Left).
* **Crucially, be specific about the exact nature of the geometric change. Do not simply state 'Modify feature.' Also take the position and its relationship to other features into account. You can output complex and multiple sentences but do not use a number of bullets here.

**Make a summary of the geometric change (Require output):**
* Output 3 distinct descriptions under a main heading: `## Geometric Summary:` for each direction.
* Based on the geometric differences identified and generated above and the images, make 3 concise and independent summaries of the geometric change.
* Under this main heading, provide **TWO distinct, concise statements**:
    * One under a sub-heading `### Left -> Right Intent:` describing 3 concise descriptions behind the change from State A (Left) to State B (Right).
    * One under a sub-heading `### Right -> Left Intent:` describing 3 concise descriptions behind the change from State B (Right) to State A (Left).
* Those descriptions should be **INDEPENDENT** from each other and **SELF-CONTAINED**.

**Infer the functional or design intent of the change (Require output):**
* Output 3 distinct descriptions under a main heading: `## Functional Intent:` for each direction.
* Based on the geometric differences identified and the summary, infer 3 likely functional or design intents behind the change.
* Under this main heading, provide **TWO distinct, concise statements**:
    * One under a sub-heading `### Left -> Right Intent:` describing 3 functional intents behind the change from State A (Left) to State B (Right).
    * One under a sub-heading `### Right -> Left Intent:` describing 3 functional intents behind the change from State B (Right) to State A (Left).
* **Crucially, be specific. Do not simply state 'Simplify ... for easier manufacturing.' or 'Improve the aesthetic ...' or 'Reduce the weight ...'. Think about the functional intent behind the change. If a change is likely made to accommodate another part, also think about and briefly mention that part and its relationship.
* Those descriptions should be separate and distinct from each other.

**Sumarize the functional intent (Require output):**
* Output 3 distinct descriptions under a main heading: `## Concise Functional Intent:` for each direction.
* Based on the images and the information above, use one sentence to summarize the function intent for each point.
* Under this main heading, provide **TWO distinct, concise statements**:
    * One under a sub-heading `### Left -> Right Intent:` describing 3 functional intents behind the change from State A (Left) to State B (Right).
    * One under a sub-heading `### Right -> Left Intent:` describing 3 functional intents behind the change from State B (Right) to State A (Left).
* Those descriptions should be separate and distinct from each other.

**ADDITIONAL NOTES** 
1. **BE PRECISE AND UNAMBIGUOUS**.
2. **ENSURE DIVERSITY IN THE INSTRUCTIONS**.
3. **Think creatively for functional/intent instructions.
4. **ACCURATELY REFLECT THE GEOMETRIC CHANGE** required for each direction (L->R and R->L).
5. **For all descriptions, act like a human user who is describing the change to another human user. Use a strong imperative verb that accurately describes the operation.
6. **Wording Constraint (Applies to both sets)**: The descriptions in each set must focus solely on describing the modification required for that specific transformation direction. **Do not** include phrases referring explicitly to the images ('in the image', 'as shown', 'on the left side', etc.). They should read as standalone instructions or change logs.
7. **Make sure that the description from both directions are meaningful without knowing the resulting image. Do not expressions like "restore/recover the feature that was present before" **
**Example format**
## Summary:
xxx
## Geometric Change:
### Left -> Right Change:
xxx
### Right -> Left Change:
xxx
## Geometric Summary:
### Left -> Right Change:
- xxx
- xxx
- xxx
### Right -> Left Change:
- xxx
- xxx
- xxx
## Functional Intent:
### Left -> Right Intent:
- xxx
- xxx
- xxx
### Right -> Left Intent:
- xxx
- xxx
- xxx
## Concise Functional Intent:
### Left -> Right Intent:
- xxx
- xxx
- xxx
### Right -> Left Intent:
- xxx
- xxx
- xxx
\end{lstlisting}

\end{document}